%% file: torino2.tex
\documentclass[12pt]{article}
\usepackage{amsmath}
\usepackage{amssymb}
\usepackage{epsfig}

\input{tfdef}

\begin{document}
\thispagestyle{empty} 
%\vspace*{-1.5cm} 
{\small hep-th/0006012}
\hfill {\small KUL-TF-2000-17} 

\begin{center}
{\large Non-commutative spaces in physics and mathematics }
\\

\vspace{1 cm} {\large  Daniela Bigatti }
\\ Institute of Theoretical Physics,
KU Leuven,
\\ B-3001 Heverlee, Belgium
\\ \vspace{2cm}

\end{center}    
Lectures given by the author at the TMR School on contemporary string 
theory and brane physics, 26 Jan--2 Feb 2000, Torino. 

\vspace{2cm}

\begin{abstract}      
The present review aims both to offer some motivations snd mathematical 
prerequisites for a study of NCG from the viewpoint of a theoretical 
physicist and to show a few applications to matrix theory and results 
obtained. 
\end{abstract}
\noindent \newpage 
{\Large Part 1: An elementary introduction to motivations and tools}
\section*{Lecture 1}
We are going to devote the first lecture to present some general remarks on 
Noncommutative geometry. We will approach it as a tool to deal with 
quotient spaces - possibly very pathological and requiring very little 
beyond the topological space structures.

\section{Introduction}{~~}
Something new is taking place in mathematics, allowing, almost for the first 
time, to study mathematical objects whose birth is not artificial but which 
used to have pathological behavior if handled with traditional mathematical 
techniques. Moreover, it is possible now to 
mimic in such context the usual tools, thus making not useless in the new 
situation all the effort 
related to the math classes. A striking feature of the new approach is the 
contrast between the simplicity and almost naivety of the questions asked and 
the qualities involved, and the 
extreme technicality of the tools needed, which has prevented potentially 
interested students 
from entering the subject. The ambition of the writer is to provide a 
(needless to say, very partial) 
introduction to noncommutative geometry without mathematical prerequisites, 
that is, aimed to 
be readable by an undergraduate student with enough patience and interest in 
the subject, 
and to give motivation for a serious study of modern mathematical techniques 
to the reader who 
wishes to enter the field as a professional. The second one will not lack 
bibliography within 
the existing literature. For the first one we feel that the revolution is 
silent, since is going on in a 
language different from his; we will try to provide a flavor of the subject, 
without hiding the 
necessity of mastering technical tools for a true comprehension. Intellectual 
honesty also 
obliges us to remark that the new ideas and techniques have very important 
elements of 
continuity with preexisting geometric and algebraic approaches, but we will 
neglect this aspect, which the reader is presumably not familiar with. 
\par Before proceeding further, let us summarize shortly our plans. 
First of all, we try to say something about the 
general philosophy 
underlying the noncommutative geometry approach. We realize, moreover, that 
the new tool, which gives interesting results also in old situations (such as 
smooth manifolds) is especially tailored to handle quotient spaces. We will 
give an example of the method in a case which has the advantages of being 
reasonably simple and reasonably typical. This will also give us the 
opportunity of discussing some rather technical mathematical tools and, 
if not follow the details, at least see some reason of why the subject is so 
linked with heavy mathematics. Hopefully the reader will then be able to 
follow discussion of physical applications, in particular those related to 
Matrix theory, some of which will be presented in the second part, 
and to the recent results in this direction. 

\section{What is geometry?} Since its very beginning, when it was almost a
surveyor's work, geometry has always  involved the study of spaces. What became
more and more abstract and refined is the  concept of (admissible) spaces, as
well as the tools for investigating them. The quest toward  abstraction and
axiomatization emerged very early and was rather advanced already at Euclid's 
times. More recent, and achieved only in modern age, were two ideas which are
going to be crucial in the following. First of all, emphasis was moved from the
nature of objects involved  (e.~g.~points or lines in Euclidean geometry) to
the relations between them; this allowed  splitting between the abstract
mathematical object, unchained from any contextual constraint,  and the model,
which can be handled, studied and worked on with ease. Moreover, it was 
realized that it is very convenient to define geometrical objects as the
characteristics which  are left invariant by some suitably defined class of
transformations of the space. Let's give the  reader an example, by giving the
definition of a topological space. Our intention is to define a  framework in
which it is sensible to speak of the notion of ``being close to'' (without the
help of  a notion of distance); we would like to require the very naive
property that ``if I take a close  neighbor of mine, and choose a close enough
neighbor of his, the last one is still my  neighbor''. To this aim, we require
first of all the space to be a set (a demand much less  obvious of how we have
been trained to assume), so that it makes sense to take subsets of  the space
and to operate on them by arbitrary union and finite intersection. Over this
space $X$  we assign a family of subsets $\cal U$, called topology, which has
to satisfy the following:  
\begin{enumerate}  \item{$ \emptyset \in {\cal U}$}
\item{$X \in {\cal U}$} \item{$\displaystyle{ V_\alpha \in {\cal U}, \alpha \in
I \Rightarrow  \bigcup_{\alpha \in I}{V_\alpha}  \in {\cal U} }$}
\item{$\displaystyle{ V_i \in {\cal U}, i=1, \ldots n \Rightarrow 
\bigcap_{i=1}^{n}{V_i} \in {\cal U} } $} \end{enumerate}  A topological space
is assigned by choosing the pair $(X, {\cal U})$; the subsets $u \in {\cal U}$ 
are also called open sets (for the topology ${\cal U}$). The axioms above can
also be stated as requiring the empty set and the whole space to be open, and
forcing arbitrary union and  finite intersection of open sets to remain open. 
\par It is clear at once that studying the topological space by handling the
family ${\cal U}$, as  the axioms above might suggest, is extremely
inconvenient, since it involves not only heavy  operations over sets, to which
we are not used, but also handling, in general, of an  enormous number of
subsets (in principle, they might be as many as ${\cal P}(X)$). This  approach
can be followed, in practice, only if we find a way of drastically truncating
the  number of open sets which are necessary for a complete characterization of
the topology.  Otherwise, it is evidently doomed to failure and we need tools
of different nature.  
\par The first observation is that a topological space is
the natural framework for defining a  continuous function. Let's then consider
functions  $ f: X \rightarrow \Bbb R$ or $f: X \rightarrow \Bbb C$ (the best
way of probing $X$ is, of course,  to make computations over the real or
complex numbers we know so well). The numerical  fields $\Bbb R$ or $\Bbb C$
have, of course, a ``natural'' topology inherited by more refined  structures.
We will say that $f$ is continuous if the counterimage of an open set of $\Bbb
R$  or $\Bbb C$ is still an open set for the topology ${\cal U}$ chosen on $X$.
(This, of course,  depends crucially on both the elements $(X, {\cal U})$ of
the topological space pair.)  
\par Let us recall, at this point, some standard definitions.  
\par A monoid is a set $M$ endowed with an associative operation,
which we denote, for  example, as $\mathbf \cdot$, ${\mathbf \cdot }: M \times
M \rightarrow M$, and containing a  so-called neutral element, which we denote,
for example, $\mathbf 1$, such that  ${\mathbf 1 \cdot} m = m {\mathbf \cdot 1}
= m \: \forall m \in M$. The standard example of a  monoid is $\Bbb N$ endowed
with addition.  
\par A group is a set $G$ again endowed with an associative
operation  ${\mathbf \cdot}: G \times G \rightarrow G$ and a neutral element
such that  ${\mathbf 1 \cdot } g = g {\mathbf \cdot 1} = m \: \forall g \in G$,
but also such that to any $g \in  G$ we may associate another $g' \in G$
(called inverse of $g$) such that $g {\mathbf  \cdot} g' =  g' {\mathbf \cdot} 
g= {\mathbf 1}$.  The standard example is $\Bbb Z$ endowed again with addition. 
\par A ring (with unity) is a set $R$ endowed with two associative operations, 
which we denote respectively as $\mathbf +$ and $\mathbf \cdot $, such that $R$
is a  commutative group with respect to $\mathbf +$, a monoid with respect to
$\mathbf \cdot $, and  enjoys distributive properties. If $R$ is a commutative
monoid for the product, it is called a  commutative ring.  
\par A field $k$ isa commutative ring with the property that $k \setminus \{ 0 
\}$ is a multiplicative group (that is, we require any nonzero element to be
invertible).  
\par A module over a ring $R$ is an abelian group endowed with a
multiplication by the  elements of the ring. (It is built in the same spirit as
a vector space, with a ring replacing  $\Bbb R$ or $\Bbb C$).  
\par An algebra over a field $k$ is a ring $A$ which is also a module over $k$ 
and enjoys a property of compatibility of the algebra product with the 
multiplication for a number (an  element of the field). (Example: the 
continuous functions $f: \Bbb R \rightarrow \Bbb R$  (resp.~ $f: \Bbb C 
\rightarrow \Bbb C$) are an algebra over $\Bbb R$ (resp.~$\Bbb C$), as we are 
about to discuss in the next few lines).   
\par It is now clear that, if we define sum and product of real (or
complex) valued functions in  the obvious way $(f {\mathbf +}g) (x) := f(x) +
g(x)$, $(f {\mathbf \cdot}g) (x) := f(x) \cdot g(x)$,  we find out that the
real (or complex) valued continuous functions form a commutative algebra. 
Since this is obtained from the addition and multiplication of $\Bbb R$ or
$\Bbb C$ only, we  remark that the same construction can be carried out for, as
an example, matrix valued continuous functions, the only difference being the
loss of commutativity.  
\par The idea of studying a topological space through
the algebra of its continuous functions is the central idea of algebraic
topology. Likewise, algebraic geometry studies characteristics  of spaces by
means, for example, of their algebra of rational functions. In this framework
it is  very natural to ask what happens if we replace the algebra of ordinary 
functions with some noncommutative analogue, both to extend the tools to new 
and previously intractable contexts (we will see soon how sometimes this 
replacement is unavoidable) and to study with more refined probes the 
``classical spaces'' (which leads sometimes to new and surprising results). 

\section{Quotient spaces} A class of typically intractable (and typically very
interesting) objects is reached by means of  a quotient space construction,
that is, by considering a set endowed with an equivalence  relation fulfilling
reflexivity, symmetry and transitivity axioms and identifying the elements
which  are equivalent with respect to the above relation. In the following we
will find particularly useful  the ``graph'' picture of an equivalence 
relation: if we consider the Cartesian product of two copies of the set, we can
assign the  equivalence relation as a subset of the Cartesian product (the one
formed by the couples  satisfying the relation). The construction gives
interesting results already if we consider a set  (possibly with an operation),
but of course is much richer if we act over a set with structure. Let's  give
some examples. First of all, let's remind the reader of the construction of
integer number  starting with the naturals.  
\par We have $(\Bbb N, +, \cdot,
<)$, that is, a set with two binary operations (by the way, both  associative
and commutative) and a relation of order. We would like to introduce the idea
of  subtraction, in spite of the notorious fact that it is not always defined.
We would like, to be more  precise, to extend $\Bbb N$ so that a subset of the
new object is isomorphic to $\Bbb N$ and  the isomorphism respects the
operations and the order relation, but which is a group with  respect to
addition. We would like, in other words, to taste the forbidden fruit 
\begin{eqnarray} \displaystyle{  m-n }  \end{eqnarray}  and, to do so, we label
it by the two integers $m$, $n$, silently meaning that  \begin{eqnarray}
\displaystyle{  m-n = m' - n'  \: \: \Longleftrightarrow \: \: (m, n) \:
{\mathrm{and}} \:  (m', n') \: {\mathrm{are \: the \: same}}   } 
\end{eqnarray}  Here arises the suggestion for an equivalence relation. Since,
though, writing the above  equality is forbidden in $\Bbb N$, we define the
equivalence relation as  \begin{eqnarray} \displaystyle{  (m,n) \sim (m', n') 
\: \Longleftrightarrow \: m + n' = m' + n }  \end{eqnarray}  and the set of
integer numbers as  \begin{eqnarray} \displaystyle{  \Bbb Z = \frac{\Bbb N
\times \Bbb N}{\sim}  }  \end{eqnarray}  where the above notation $/ \sim$
means identification with respect to $\sim$.  
\par This construction fulfils
all requirements. We beg pardon from the reader  for being so pedantic; we just
would like to have a trivial example as a guide  for more abstract contexts. 
\par Let's see instead which dangers may occur if we have to do with a more 
refined space, say, a compact topological space. We say that a topological 
space with a given topology is compact if any open covering  of the space
(i.~e.~any family of open sets whose union is the space itself)  admits a
finite subcovering (i.~e.~a finite subfamily of the above which is  still a
covering). We say thet a topological space $(X, {\cal U})$ is locally  compact
if for all $x \in X$ and for all $U \in {\cal U}$, $x \in U$, there  exist a
compact set $W$ such that $W \subset U$.  It is, of course, useful to refer to
a compactness notion also  when we have more structures than just the one of a 
topological space (for example, differentiability).  
\par Let's now consider
the flat square torus, that is, $[0,1] \times [0,1]$ with the opposite sides 
ordinately glued. This is clearly a well-behaved space and, undoubtedly, a
compact one.  Let's introduce an equivalence relation which identifies the
points of the lines parallel  to $y= \sqrt{2} x$, that is, we ``foliate'' the
space into leaves parametrized by the intercept. Since  $\sqrt{2}$ is
irrational, though, any such leaf fills the torus in a dense way (that is,
given a leaf  and a point of the torus, the leaf is found to be arbitrarily
close to the point). If we try to study the  quotient space and to introduce in
it a topology, we will find that ``anything is close to anything'',  that is,
the only possible topology contains as open sets only the whole space and the
empty  set. It is hopeless to try to give the quotient space an interesting
topology  based on our notion of ``neighborhood'' of the parent space.  
\par It is, in particular, hopeless for all practical purposes to give the 
space the standard notion of topology inherited by  the quotient operation, 
which we shortly describe. If we have a space $A \equiv B / \sim$,  there is a 
natural projection map  \begin{eqnarray}  \begin{array}{c}  p \: : 
\:\:\:\:\:\:\:  B \longrightarrow A  \\  p \: : \:\:\:\:\:\:\:  x \longmapsto 
[x]  \end{array}
\end{eqnarray} 
which sends $x \in B$ in its equivalence class. The inherited
topology on $A$  would be the one  whose open sets are the sets whose
counterimages are open sets in $B$.  
\par We want an interesting topology,
richer than $\{  \emptyset , X \}$ ,  and, moreover, we would like the
topological space so obtained to enjoy local  compactness. The reason why we
make the effort is that the dull topology  $\{  \emptyset , X \}$ treats the
space, from the point of view of continuous  functions which will be our probe,
as the space consisting of only one point;  all the possible subtleties of our
environment will be lost. Local  compactness is a slightly more technical tool,
but we can imagine, both from  the physical and the mathematical point of view,
why it is so useful.  Each time we have a nontrivial bundle (and we will have
plenty of them in  the following) we usually define them not globally, but on
neighbourhoods.  Since these ``patches'' will in general intersect, we need a
machinery  to enforce agreement of alternative descriptions. Local compactness
(and  similar tools) ensure us that ``the number of possible alternative 
descriptions will never get out of control''. We shall see how to achieve the 
notable result of introducing in ``weird'' spaces a rich enough and  even
locally compact topology. An example of the process is presented in the  next
section. 

\section{A typical example: the space of Penrose tilings} We are going to
discuss a situation which embodies most of the characteristic features both of 
the problems which noncommutative geometry makes tractable and of the procedure
which  allows their handling. Notably enough, such a space has recently been  
given hints of physical relevance: see \cite{daniela}. 
\par Penrose was able to build tilings of the
plane having a 5-fold symmetry axis; this is not  possible by means of periodic
tilings with all equal tiles, as it is known  since a long time.
They are composed (see figure 1) of two types of tiles:  ``darts'' and 
``kites'', with
the condition that every vertex has matching  colors. A striking characteristic
of the Penrose tilings is that any finite  pattern occurs (infinitely many
times, by the way) in any other Penrose  tiling. So, if we call identical two
tilings which are  carried into each other by an isometry of the plane (this is
a sensible  definition since none of the tilings is periodic), it is never
possible to  decide locally which tiling is which. We will give, first of all,
arguments in  favour of the existence of really different tilings and, second,
methods  to actually discriminate among them. In order to gain some mental
picture,  we anticipate that it turns out that the notion of average number of
darts  (resp.~kites) per unit area is meaningful, and the ratio of these two 
averages is the golden ratio; moreover, the distinct Penrose tilings are  an
uncountable infinity.  
\par There is a very important result which allows us to
parametrize such  tilings with the set $K$  of infinite sequence of zeros and
ones satisfying a consistency condition:  \begin{eqnarray} \label{*}
\displaystyle{  K \equiv \left\{  (z_n), \: n \in {\Bbb {N}}, \: z_n \in 
\{ 0, 1
\}, \: \: (z_n =1) \Rightarrow (z_{n+1} =0)  \right\} }  \end{eqnarray}  This
point is crucial, but we feel unnecessary to write down the proof, which the
reader can find,  for example, in [1],  
%\cite{Connes},  
cap.~2 appendix D. The
reader is advised to go through the  graphical details of the construction, in
order to become convinced of how two different  sequences $z$, $z'$ lead to the
same tiling if and only if they are definitely identical\footnote{A  warning
for the English speaking readers: ``definitely identical'' is the expression we
are  about to define, it does not mean ``really equal''.}:  \begin{eqnarray}
\displaystyle{  z \sim z' \: \: \: \Longleftrightarrow \: \: \: \exists \: n
\in {{\Bbb {N}}} \:  {\mathrm{ s.~t.}~~}z_j = z'_j \: \:  \forall \: j \geq n  } 
\end{eqnarray}  The space $K$ is compact and is actually isomorphic to a Cantor
set. (Actually  this story should be told properly, paying due attention to the
topology  we put over $K$, but we are not going to do it here). To the reader
who is not  familiar with this construction, we remind that a Cantor set is
built by taking the interval $[0, 1]$,  dividing it in three parts and throwing
away the inner one; then we iterate the procedure,  throwing away in the second
step the intervals $(\frac{1}{9}, \frac{2}{9})$  and $(\frac{7}{9},
\frac{8}{9})$, and so on (see figure 2), till we are left  with a ``dust of
dots'', a compact set totally disconnected (this means that  none of  its
points has a connected neighborhood; in its turn, a connected neighborhood is
one which  cannot be written as union of two open disjoint empty sets; it is
apparent in our example that  any neighborhood of any point of $K$ can be
written as union of two disjoint open sets), but without isolated points (any
neighborhood of a  point of $K$ contains other points of $K$). (It is, by the
way, proved that, up to homeomorphisms,  any set with the above characteristics
is a Cantor set.)  
\par To realize that our space of numerical sequences has
something to do  with a Cantor set, we suggest another construction for $K$,
the so-called  Smale horseshoe  construction. Let's consider a transformation
$g$ acting over ${\cal S}= [0,1] \times [0,1]$ and  transforming the square in
a horseshoe (see figure 3):  \begin{eqnarray} \displaystyle{  \bigcap_{n \in
\Bbb N}{g^n ({\cal S})} \:  = \: [0, 1] \times  {\mathrm{ Cantor \: set }}  } 
\end{eqnarray}  The Smale construction allows to construct a biunivocal mapping
between the  points of the Cantor set and the infinite sequences of zeros and
ones. This is  achieved by proving, thanks to the ``baking'' properties of the 
transformation, which ``stretches'', ``squeezes'' and ``folds'' simultaneously,
that ${\cal S} \cap g({\cal S})$ contains two connected components, $I_0$ and 
$I_1$, s.~t.~$g({\cal S}) \cap {\cal S} = I_0 \cup I_1$. It is thus very easy
to  define the mapping between $x \in K$ and $(\zeta_n)_{n \in \Bbb N}$ with
values in $\{ 0, 1 \}$:  we have only to define $\zeta_n =0$ (resp.~$=1$) if
$g^n(x) \in I_0$ (resp.~$I_1$). Moreover, we  point out that substituting $x
\in K$ with $g(x) \in K$ corresponds to a left shift of the sequence 
$(\zeta_n)$. We can imagine a particular horseshoe transformation satisfying, 
in addition, the  consistency condition of (\ref{*}). \footnote{The readers
skilled in advanced  mechanics may have some doubts that it is possible to
satisfy the  consistency condition of (\ref{*}) and at the same time not to
spoil the  interpretation of $I_0$ and $I_1$ and their properties
(connectedness in  the first place). The paragraph just meant to give the
reader an idea of how  to relate the manipulation on $[0,1]$ which lead to
construction of $K$ and  the parallel building of a space of binary sequences.
This does not imply we  are literaly mimicking the Smale construction. }  
\par Summarizing, we have a compact space $K$, homeomorphic to a Cantor set: 
\begin{eqnarray} \displaystyle{  K \equiv \left\{  (z_n), \: n \in {\Bbb {N}}, 
\:
z_n \in \{ 0, 1 \}, \: \: (z_n =1) \Rightarrow (z_{n+1} =0)  \right\} } 
\end{eqnarray}  and a relation of equivalence $\cal R$ \begin{eqnarray}
\displaystyle{  z \sim z' \: \: \: \Longleftrightarrow \: \: \: \exists \: n
\in {\Bbb{ N}} \: {\mathrm s.~t.~~}z_j = z'_j \: \:  \forall \: j \geq n  } 
\end{eqnarray}  The space $X$ of Penrose tilings is the space  \begin{eqnarray}
\displaystyle{  X = K / {\cal R} }  \end{eqnarray}  
\par We have already
pointed out how, at first sight, there exist only one  Penrose
tiling\footnote{The careful reader might be confused by this remark:  obviously
there is more than one (and actually an awful lot),  since the equivalence
classes contain a denumerable infinity of elements, while $K$ has the
cardinality of continuum. What we want to do is not a counting,  but to make
sure that the tilings are different in an {\it interesting} sense. Let's
explain a little more this slippery issue. The ``intuitive'' approach  to
discrimination would be to give properties (such as relative frequency of 
appearance for darts and kites) over ``regions of bigger and bigger radius''. 
We already know it won't work, and this is a typical feature of noncommutative 
sets: a denumerable set of comparisons is not enough to make sure that two 
tilings are different. In such a situation, with no concrete possibility of 
operationally distinguishing the elements of $X$, we may be tempted to give up 
and treat noncommutative spaces in the same way as non measurable functions: 
we know they exist and are actually the majority, but we will never write one 
of such, and so we leave the axiom of choice to be studied by set theorists. 
Our case is a bit different and cannot be ignored light-heartily exactly 
because of the existence of abelian groups labeling with different numerical 
answers the ``subtle'' differences asking for such attentive mathematical 
description. It should anyhow be stressed that we have to be really careful 
while using concepts like cardinality in a non commutative space 
(cfr.~[1]).}, and the  
%%%%(cfr.~\cite{Connes}). }, and the  
idea of discriminating among
them by means of the algebra of continuous functions is hopeless.  The path we
will follow is to show (1) that the attempt of distinguishing the tilings by
means of  an algebra of operator-valued functions is successful, and (2) that
it is actually sensible to say that there are different Penrose tilings, since
topological invariants can be built and used to label  the tilings. The process
requires, of course, some mathematical work, which we postpone to the next two
sections. For point (1) the mathematical prerequisites are completely standard: 
essentially  knowledge of Hilbert spaces and of the classical spaces of
functional analysis (such as $l^2$).  For (2) we need some K-theory notions,
which we will try to summarize in the next section. 

\section{What is K-theory?} This section is going to be rather technical and
not really necessary in order to follow the  sequence of steps leading to a
treatment of the diseases of the previous section. The reasons for  including
it in this elementary review are to explain what the K-groups used in the
construction of  ``labels'' for Penrose tilings and similar ill-behaved spaces
are, and to satisfy the curiosity of the  reader of [2]  
%\cite{CDS}  
who is wondering about the surprise of even K-theory groups instead of odd 
ones and about the relation between K-theory and cyclic cohomology.  
\par Since the
procedure is going to be very technical, we feel the need to summarize both the 
succession of steps and the purposes. We are going to build a new cohomology
theory which  enjoys some technical advantages and a remarkable property called
Bott periodicity: the  relevant K-cohomology groups are only two, $K^0$ and
$K^1$, since all the odd order groups  are isomorphic to $K^1$ and all the even
order ones to $K^0$. There is actually a big question:  these K-theory groups
are very appealing, but it's not obvious that they are computable.  The answer
is fully beyond the size of this review, anyway sometimes they are, and other
times  one can compute cyclic cohomology group, which give ``similar''
information. There is, besides,  a very non trivial extension of these tools to
the noncommutative context.  
\par Anyway, let's try, as promised, to summarize
the (steep) steps of the construction. First of  all, the appropriate language
is the one of category theory, that is, a mathematical monster  consisting of
``objects'' (for example vector spaces, topological spaces,  groups, etc.), of
``morphisms'', that is, maps between the above objects (for  example, linear
maps, continuous maps, group morphisms, etc.),  and some appropriate rule about 
compositions. We need this language since we want to explain the meaning of
some expressions  (for example, the difference between topological and
algebraic K-theory) which the reader  might be curious about (since has not
quit reading up to now); as we will point out, though, it is  indispensable
only in some respect and it is useful, but not necessary, for the rest.  
\par Let us take a so-called additive category, that is, one in which 
``summing two
objects'' is  meaningful (for example the one of finite dimensional vector
spaces: we can define $V \oplus  U$). If we take an object of such, say $V$,
and consider ${\cal I}(V)$, the class of isomorphisms  of $V$, we see that we
can define a sum for such class of isomorphisms: ${\cal I}(U) + {\cal I}(V)  :=
{\cal I}(U \oplus V)$. Neglecting all technical details, we just point out that
this sum induces a  monoid structure, but not a group. We would like a group
instead, since we don't like  cohomology monoids. We will have to discuss how
to build a group (in some sense, the ``most  similar'' group) out of a monoid;
we will see shortly that this operation is identical in spirit  to building
$\Bbb Z$ out of $\Bbb N$. What is born is a group, called the Groethendieck
group  of the category $\cal C$, which is the starting point of the
construction.  This group will become the order zero cohomology group. For our
purposes,  we are going to choose as $\cal C$ the category of fiber bundles
over a  compact topological space. (Beware: the reader who is interested in 
understanding the treatment of the foliation of the torus when the leaf is  non
compact must remember that the tools are not yet being extended to the 
noncommutative case). The next step will be to define an extension of the 
above group, which will be defined not for a category, but for a  so-called
functor between two categories: $\varphi : {\cal C} \rightarrow  {\cal C}'$.
(Slogan: a functor is more or less like a function, but acts  between
categories.) This allows to build the $K^{-1}$ group, which leads  to all the
$K^{-n}$ groups with $n > 0$. To build the $K^n$ groups is much more difficult,
and  actually one proceeds by proving the Bott periodicity theorem, so that we
are guaranteed that  explicit construction is not necessary.  
\par We are not
at all trying to provide the reader with a description of the procedure, but we
are trying to give an explanation of what algebraic K-theory and topological
K-theory are, and to  sketch what is going on in the case of fiber bundles,
which will be relevant in the following. We  will find out that topological
K-theory is a tool for studying topological spaces, that algebraic  K-theory is
instead a tool for studying rings,  and that there is a relation between the
two when the algebra is, say, an algebra of functions  over the topological
space. The link is given by the Serre-Swan theorem, which  gives a canonical
correspondence between the vector bundles over the  topological space and the
projective modules of finite type over the algebra  of continuous functions
over it (a module is like a vector space, but instead  of having a field over
it, it has got a ring (commutative and with unit);  typically, a vector space
is a vector space over $\Bbb R$ or $\Bbb C$, that  is, we can multiply a vector
by a number and things like that; a module can be  thought of, instead, as
something similar, but with linearity  over a ring of function). See also note
4.  
\par At this point, we also want to tell the reader that the abstract 
(categorial) construction we have chosen to present is motivated by the need of
discussing  algebraic K-theory; if we study only algebraic K-theory for the
algebra of continuous functions  over a topological space, this is unnecessary,
but in a general framework we have an abstract  algebra, not connected with a
concrete topological space, and it is necessary to construct a  ``mimicked''
one by means of a tool which needs to be able to transfer information from the
world  of topological spaces and continuous functions to the world of algebras.
In this framework,  by the way, topological K-theory is just going to be a very
particular case; namely, a case in  which the, let's say, topological space has
actually a meaning.  
\par Let's first of all see how to invent a group out of a
monoid. Be given an  abelian monoid $M$, we want to construct $S(M)$, an
abelian group, and $s$,  a map $M \rightarrow S(M)$ respecting the monoidal
structures, such to enforce  the following request: given another abelian group
$G$ and given $f: M  \rightarrow G$, homomorphism respecting the monoidal
structures, we can find  a unique $\tilde{f}: S(M) \rightarrow G$, homomorphism
respecting the group  structures, such that $f(m)= \tilde{f}(s(m)) \: \:
\forall \: m \in M$, that is, $f$ can be reconstructed by acting with
$\tilde{f} (s( \:))$. To say it  graphically  \begin{eqnarray}  \displaystyle{ 
\begin{array}{c} \: \: \: M \: \: {\stackrel{s}{\longrightarrow}} \: \:  S(M)
\\  f  \searrow  \: \: \: \: \: \:   \swarrow \tilde{f} \\  G  \end{array}  } 
\end{eqnarray}  it is the same to take the ``short'' or the ``long'' path to go
from $M$ to any abelian group $G$,  where we intend the arrows to preserve ``as
much structure as they can''.  
\par Let's give the reader a couple of examples
of how one builds actually $s$ and $S(M)$ out  of $M$. Obviously, since the
solution of the problem is unique up to  isomorphisms, all such constructions
are equivalent up to isomorphisms too.  
\par Let's take the cartesian product
$M \times M$ and quotient it by means of the following  equivalence relation: 
\begin{eqnarray} \displaystyle{  (m, n) \sim (m', n') \: \: 
{\stackrel{def}{\Longleftrightarrow}} \: \: \exists \: p, q \: \:  {\mathrm{
such \: that}}  \: \:  m+ n'+p = n +m'+q  }  \end{eqnarray}  The quotient
monoid turns out actually to be a group ; to the element $m$ of the monoid one 
associates the element of the group $s(m) = [(m, 0)]$ (where the square
brackets denote the equivalence class).  
\par An equivalent construction is to
consider $M \times M$ quotiented by the equivalence  relation  \begin{eqnarray}
\displaystyle{  (m, n) \sim (m', n') \: \: 
{\stackrel{def}{\Longleftrightarrow}} \: \: \exists \: p, q \: \:  {\mathrm{
such \: that}}  \: \:  (m, n) + (p, p) = (m', n') + (q, q)  }  \end{eqnarray} 
in which case $s(m) = [(m, 0)]$ again. (A remark for the careful reader: not
necessarily the  $s$ transformation is injective.)  
\par Let's show some
examples of the construction. One of these has already be worked out,  and the
reader can translate step by step between the two languages:  \begin{itemize} 
\item{$M = {{\Bbb{ N}}} \: \: \: \: \: \: \: \: \: \: S(M) \approx {\Bbb{ Z}}$ }
\end{itemize} Another example will be comprehensible if we recall the
construction of rational numbers by  means of equivalence classes of fractions: 
\begin{itemize}  \item{$M= {\Bbb{ Z}} \setminus \{ 0 \}  \: \: \: \: \: \: \: 
\: \:
\: S(M) \approx {\Bbb {Q}} \setminus \{  0 \}$ } \end{itemize} And now a little
surprise:  \begin{itemize}  \item{Let $M$ be an abelian monoid with ${\mathbf
+}$ operation, with an element denoted  ${\mathbf \infty}$ such that $m
{\mathbf + \infty} = {\mathbf  \infty} $ (or, which is the same,  an abelian
monoid with a ${\mathbf \cdot}$ operation and an element ${\mathbf 0}$ such
that ${\mathbf m \cdot 0} = {\mathbf 0}  \: \: \forall m \in M$). Then $S(M)
\approx 0$. This happens,  for example, to $\Bbb Z$ endowed with multiplication
or to $\Bbb R \cup  \{ \infty \}$ endowed with addition. } \end{itemize}  Maybe
the reader would like to know how it is so. Let's give an argument for, as an
example,  $\Bbb N \cup \{ \infty \}$ endowed with addition. The equivalence
classes of finite numbers are  the usual ``diagonal set of points'' (see figure
4), while $[(\infty, 0)] = \{ (\infty, p), p \in  \Bbb N \}$. Since ``all the
lines will intersect in $(\infty, \infty)$'', there is only one element of the 
group, namely $[(0, 0)]$.  
\par We are now going to discuss the example of
symmetrization of a monoid which interests us  most. We will take a category
${\cal C}$ where summing two objects is meaningful (such as the  one of vector
spaces, $V \oplus U$ is meaningful). We aim to the category of vector bundles, 
as we have already suggested, since this is the heart of the interplay between
topological  spaces and rings of functions. As we already said, if we consider
${\cal I}(E)$, the class of  isomorphisms of $E$ (where $E$ is an object of the
additive category), we can define a sum:  ${\cal I}(E) + {\cal I}(F) := {\cal
I}(E \oplus F)$, and thus induce a structure of abelian monoid.  There are nice
properties: ${\cal I}(E \oplus F)$ only depends on ${\cal I}(E)$ and ${\cal
I}(F)$  and $E \oplus (F \oplus G) \approx (E \oplus F) \oplus G$, $E \oplus F
\approx F \oplus E$,  $E \oplus 0 \approx E$. Let's denote $I$ the set of the
isomorphism classes ${\cal I}(E)$. The  abelian group $S(I) \equiv K ({\cal
C})$ is called the Groethendieck group of the category  $\cal C$.  
\par Let's
give the reader some example of such groups. The first two will be very simple
ones,  which are somehow already known to the reader; the other two, instead,
will be relevant in the  following.  \begin{itemize}  \item{Let $\cal C$ be the
category whose objects are the finite dimensional vector spaces and  whose
morphisms are linear maps. We know the notion of dimension of a vector space 
(``a vector space is not much more than ${\Bbb{ R}}^n$ or ${\Bbb{ C}}^n$''), 
and this
allows us to say  that $I \approx \Bbb N$. Then, the first example of
symmetrization of a monoid tells us that  $K({\cal C}) \approx \Bbb Z$. }
\item{If we consider, instead, the category of all vector spaces (regardless of
the finite  dimensionality) it turns out $K({\cal C}) \approx 0$. This is
because we are in the situation of the  third example: if we choose $\tau : E
\longmapsto E \oplus E \oplus E \oplus \ldots$, we have  $s(\tau(E)) + s(E) =
s(\tau(E))$.  } \item{Let $R$ be a ring with an unit. As we construct a vector
space of  finite dimension over a field $k$ (typically $\Bbb R$ or $\Bbb C$) 
by requiring that there is a commutative sum for the elements of the vector 
space, which gives a commutative group structure, and there is another 
operation $k \times V \rightarrow V$, that is, multiplying a vector by a 
scalar, with a certain number of properties, in the same way we can build a 
projective module of finite type\footnote{The definition of finite type 
projective module is not  irrelevant, since vector bundles over a compact space
$X$ can be identified  with projective modules of finite type over $C(X)$, the
algebra of complex  valued continuous functions over $X$; since this statement
is almost the  starting point of noncommutative geometry, we would like to
exploit this  footnote to give more precisely this notion.   \begin{itemize} 
\item{Let $S$ be a set; the $R$-module $F$ is said to be free generated by $S$ 
if, given an injective map $i: S \rightarrow F$ which defines a family $f_i$ 
of elements of $F$ indexed by $i \in S$ (the $f_i$ are called generators), 
and, given another $R$-module $A$, where another map $h: S \rightarrow A$ 
defines $a_i \in A$, $i \in S$, $\exists ! \: k: F \rightarrow A$ such that 
$k(f_i) = a_i$. (Vector space analogy: $S= \{ 1, \ldots, {\mathrm{ dim}} \, V 
\}$; a linear mapping is uniquely reconstructed if we know where the 
generators of the domain space are landed). }  \item{A module is said to be
generated by a set $S$ if all its elements can be  written as $\displaystyle{
\sum_{s \in S}{r_s \: s}  \:\:\: r_s \in R  }$   where only a finite number of
$r_s$ is nonzero. }  \item{It is said to be finitely generated or of finite
type if it is generated by a finite set. }  \item{An $R$-module is said to be
projective if and only if it is a direct summand of free $R$-modules. } 
\end{itemize}  }  over a ring $\cal R$ (possibly a ring of functions); notice
that if $\cal R$  is a ring of functions the notion of linearity ($\cal
R$-linearity) is much  more problematic, since it involves ``taking inside and
outside of brackets''  not numbers but functions. Let's consider, then, the
category which has as  objects the finite type projective $R$-modules (see
again footnote 4) and as  morphisms the $R$-linear maps. The Groethendieck
group of such is usually  denoted $K(R)$. The aim of algebraic K-theory is to 
compute $K(R)$ for (interesting) rings $R$. } \item{Let $\cal C$ be the
category of the vector bundles over a compact topological space  $X$. Its
Groethendieck group is usually denoted $K(X)$. The aim of topological K-theory
is to  compute $K(X)$ for (interesting) topological spaces $X$. } \end{itemize}
\par The relation between topological and algebraic K-theory is given by the
observation that  the respective K-theory groups are isomorphic provided one
proves the equivalence of the  related categories (theorem of Serre-Swan). This
can be done and the crucial ingredient in  the proof is the so-called section
functor, which states the fact that the set of the continuous  sections of a
vector bundle is a $R$-module and some further ``good behavior'' properties. 
\par To proceed along the path outlined, we want to make some further
observations. First of  all, we did not really need to own a notion of
additivity inside the category; it would be enough  to have a composition law
$\bot$ satisfying the same ``nice properties'': $E \bot (F \bot G)  \approx (E
\bot F) \bot G$, $E \bot F \approx F \bot E$, $E \bot 0 \approx E$. This allows
us to  study in this framework a very significant example: the real finite
dimensional fiber bundles over  a compact topological space $X$ endowed with a
positive definite quadratic form.  It turns out  $[E] \approx [F]$ if and only
if $ \exists \: G$ such that $E \bot G \approx F \bot G$ (actually, $G$  may be
chosen to be a trivial vector bundle of suitable rank $n \in \Bbb N$ over X);
in this case  one says $E$ and $F$ to be stably isomorphic. One finds that to
any $E$ one can associate  $[E] \in K^0(X)$, which depends only on the class of
stable isomorphisms of $E$. Such classes  $[E]$ generate the group $K^0(X)$.
The group $K^1$ is more interesting, since $K^1(X) =  \pi_0 (GL^{\infty})$; it
is the group of connected components of $\displaystyle{GL^{\infty} \equiv 
\cup_{n \geq 1} GL^n (A)}$, where $GL^n$ are $ n \times n$ invertible matrices
whose elements  take values in $A=C(X)$.  
\par Going back to the general
context, another thing to be noticed is that $K$ is a  contravariant functor
(that is, speaking very loosely, ``the analogous of a map between  categories,
but reversed; it would be the generalization of a map from the ``target'' 
category to the ``domain'' one'') from the category $\{$compact topological
spaces and  continuous applications$\}$ to the one $\{\Bbb Z/2$-graded abelian
groups$\}$ (the  gradation comes from $K = K^0 \oplus K^1$).  
\par Contravariancy is very important: if one ``squeezes'' the topological 
space, either to a  point or a subset of his, by means of an equivalence 
relation, one obtains a well-defined map  between abelian groups (notice how 
the arrows should be oriented for this to work).  
\par The steps which follow are very
technical and we don't want even to outline them; let's  only say that
contravariancy is crucial and that one needs to build a product structure 
which is able to multiply elements of K-groups associated to (different)
topological spaces.  One thus proves the Bott periodicity theorem and
``builds'' in a very abstract way all the K-theory  groups.   
\par Let's say
something (addressing the reader wishing a textbook to  [1]) about the 
%\cite{Connes}) about the 
very  important questions of computability of
K-theory groups and of extension of the tools to the non  commutative case.
Let's take $A=C(X)$, $X$ compact space. In this (commutative) case, we  have a
relation between the K-theory of $X$ and its usual (de Rham) cohomology. This
tool  (the Chern character) can be explicitly computed when $X$ is a smooth
manifold. The columns  on which this results stands are:  \begin{itemize} 
\item{there is an isomorphism $K_0({\cal A})   \simeq K_0 (A)$, where  ${\cal
A}= C^{\infty}(X)$ is dense in $A$; }  \item{given $E$, a smooth vector bundle
over $X$, we associate to it an element of the  cohomology, $ch(E)$, which can
be represented as the differential form $ch(E)=  trace(exp(\Delta^2 / 2 \pi
i))$ for any connection $\Delta$ over $E$; } \item{if we take a continuous
linear form $C$ over the vector space of the smooth differential  forms over
the manifold $X$, one can ``couple'' it with the differential form above:
$\langle  C, ch(E) \rangle \equiv \varphi_C (E)$; thus $\varphi_C$ is a map
$\varphi_C : K^*(X)  \rightarrow \Bbb C$.  } \end{itemize}  The procedure has
thus managed to give us numerical invariants in K-theory, whose knowledge  may
replace the actual study of $ch(E)$.  
\par To allow extension to the
noncommutative case, we need two big steps.  \begin{itemize}  \item{We need to
define the non commutative analogue of the de Rham cohomology, which will  be
called cyclic cohomology; this step is algebraic and needs figuring out an
algebra $\cal A$  which will play the role of $C^\infty (X)$. }  \item{The
second crucial step is analytic: given $\cal A$, non commutative algebra which
is a  dense subalgebra of a $C^*$-algebra\footnote{ The general proofs of
K-theory are based on  Banach algebra conditions, but there are crucial
differences between the K-theory of  $C^*$-algebras and the one of generic
involutory Banach algebras. Here we do need  $C^*$-algebras. } $A$, let's have
a cyclic cocycle over $\cal A$ (with suitable conditions).  We need to extend
the numerical invariants of K-theory; at this stage, we have actually 
$\varphi'_C : K_0 ({\cal A}) \rightarrow \Bbb C$ and we need $\varphi'_C : K_0
(A) \rightarrow  \Bbb C$.  } \end{itemize}  
\par We have thus outlined the path
towards the construction of a very powerful tool, which  has been able to
handle many hard problems in mathematics. To the non mathematical  audience, we
now would like to remark that it allowed comprehension of (algebraic)  relation
between differential geometry and measure theory. For example, in the context
of a  foliate manifold (when usual measure theory may be helpless), it was
possible to recover the naive interpretation of a bundle of leaves over the
ill-behaved space of leaves, provided everything is correctly reformulated in
the noncommutative framework. 

\section{The enigma of Penrose tilings and its solution}  We have discussed in
section 4 an ill-behaved space, obtained by a compact one by means of  a
quotient operation, and have seen how it was hopeless to resolve its structure
by means of  the algebra of continuous functions. As promised, now we move on
towards showing that there  are actually different Penrose tilings, since there
are topological invariants labeled by integers:  the actual picture will be
that it is true that any finite patch will appear (infinitely many times)  in
any Penrose tilings, but the limit, for bigger and bigger regions of the
tiling, of the relative  frequency of appearance of patterns can be different.
Moreover, the job is done by studying  operator-valued functions over the
ill-behaved space $X$. The section will require some  (standard) knowledge of
Hilbert spaces and linear operators over them.  
\par Let us show, first of all,
what the $C^*$-algebra of the space $X = K / {\cal R}$ of Penrose  tilings
looks like. A generic element $a$ of the $C^*$-algebra $A$ is given by a matrix 
$\displaystyle{(a)_{z, z'}}$ indexed by pairs $(z, z') \in  {\cal R} \subset K
\times K$.  The product of the algebra is the matrix product:
$\displaystyle{(ab)_{z, z'} = \sum_{z''}{ a_{z, z''} b_{z'', z'}\,}}$.  
\par To
each $x \in X$ one can associate a denumerable subset of $K$ (the sequences of 
numbers which are definitely equal), and, thus, a Hilbert space $l^2_x$ (of
which the above  sequences give an orthonormal basis). Any $a \in A$ defines an
operator of this Hilbert space  $l^2_x$:  \begin{eqnarray} \displaystyle{ 
\left(  (a(x)  \xi   \right)_z = \sum_{z'}{a_{z, z'} \xi_{z'}}   \: \: \: \: \:
\: \: \forall \xi \in l^2_x }  \end{eqnarray}  that is, $a(x) \in L(l^2_x) \:
\: \forall x \in X$. This, by the way, defines a $C^*$-algebra, since 
$||a(x)||$ is finite and independent of $x$.  
\par Notice that $\cal R$ can be
endowed with a locally compact topology. We can, actually,  see $\cal R$ as
$\displaystyle{ \bigcup_{n \in \Bbb N}{{\cal R}_n}}$, where the relations 
${\cal R}_n$ are defined as  \begin{eqnarray} \displaystyle{  {\cal R}_n =
\left\{  (z, z') \: : \: \: \: z_j = z'_j \: \: \: \: \forall \: j \geq n 
\right\} }  \end{eqnarray}  that is, we consider the pairs whose sequences
match since the first, second, ... $n$-th figure.  
\par It is clear how to
extract from any covering a finite subcovering (for example, if we deal with 
neighborhoods of a point, ``take the smaller $n$, that is, the coarser 
cell''). It is also clear that  this topology is not equal to the topology
which $\cal R$ would inherit as a subset of $K \times  K$ (that is, the
topology which says ``two couples are near if the first elements and the second 
elements of the couples are respectively near'').  
\par We don't want here to
give an explicit construction of the $C^*$-algebra $A$ (which the  reader can
found discussed in detail in [1],  
%\cite{Connes},  
II.3), but only to stress
again the two  important points:  \begin{enumerate} \item{the $C^*$-algebra $A$
is rich and interesting;} \item{it has invariants labeled by integers.}
\end{enumerate} Let's just summarize the results obtained in this direction.
$A$ turns out to be the inductive limit  (this term should not frighten the
reader: it just mean a limit of bigger and bigger matrices,  boxed one into the
previous) of the finite dimensional algebras $\displaystyle{  A_n = M_{k_n}
({\Bbb {C}}) \oplus M_{k_n'} ({\Bbb {C}}) }$. That is, the $A_n$ are direct sums of two
matrix algebras  of respective dimension $k_n$ and $k_{n'}$; in their turn,
$k_n$ and $k_{n'}$ are natural  numbers obtained by the following steps: (1)
truncate the binary sequence representation of the  Cantor set $K$ to the sets
$K_n$ of finite sequences of length $n+1$ satisfying the  consistency rule, and
(2) set equal to $k_n$ (resp.~$k_{n'}$) the number of finite sequences  of
$K_n$ which end with zero (resp.~one). It is known how to calculate invariants
for such an  algebra $A$, which are, by the way, the $K_0$ group and its
symmetrization (this remark is  aimed to the readers who read the previous
section or who are familiar with K-theory). The  result is  \begin{eqnarray}
\displaystyle{  K_0 (A) = {\Bbb {Z}}^2 }  \end{eqnarray}  \begin{eqnarray}
\displaystyle{  K_0^+ (A) = \left\{ (a, b) \in {\Bbb{ Z}}^2 \: : \: \left( 
\frac{1
+ \sqrt{5}}{2} a + b \geq  0 \right)  \right\}  }  \end{eqnarray}  
\par One can
thus achieve the construction of numerical invariants (which can be interpreted
as  relative frequency of appearing for finite patterns); this justifies all
the effort put in the direction of  resolving the structure, at first sight
trivial, of the non-pointlike set $X$.  
%
% \begin{eqnarray} \displaystyle{ 
% }  \end{eqnarray} 
\newpage 
\noindent 
{\Large Part 2: Some applications to Matrix theory}  

\section*{Lecture 2}
The noncommutative torus: a very simple example of a non commutative 
geometry, which also allow us to figure out relations between the 
deformation viewpoint and the quotient space viewpoint. 

\section*{Introduction}
In this lecture I will present a very simple example of noncommutative geometry, the 
noncommutative torus, which moreover we will study mainly in the case of rational foliation 
(that is, a geometrically trivial framework). This example will allow us to clarify the 
relation between the ``deformation'' aspect which is intrinsic to noncommutativity and 
the insurgence of a magnetic field (or charge) which on the other hand is a reason of 
interest from a more ``physical'' viewpoint. We will start by discussing a little the 
sentences that one often encounters about ``non commutative manifolds'' that are endowed 
with non commuting coordinates and show how they actually say no more and no less that 
the algebra of ``functions'' which describes the space is actually an algebra of 
(non commuting) operators. We will then go through the construction of the paper 
hep-th/9804120 which links an abelian gauge theory over a noncommutative torus 
with a nonabelian gauge theory in presence of magnetic monopoles over an ordinary 
torus, and relates as well the original parameters for the noncommutative torus 
with parameters describing the gauge group and the magnetic field in the rephrased theory. 
This means that in such a framework the operation of deforming a commutative theory into 
a noncommutative one just means acting over the magnetic boundary conditions of the theory 
and the rank of the gauge group - the noncommutative description is more convenient in 
some sense and as we will see clarifies for us a relation between different SYM which 
is not at all evident from the traditional point of view, but it is actually a rephrasing. 
So we should discuss in further detail how far we expect NCG in M theory to be just a 
rephrasing of known physics, and we will talk again of this aspect in the next lecture.

\section{The NC torus}  

Let us consider the noncommutative torus whose characterization has been given in the 
previous lecture 

\begin{eqnarray}
U_i * U_j = e^{2 \pi i \theta_{ij}} U_j * U_i
\end{eqnarray}

\begin{eqnarray}
U^i \equiv e^{i x_i}
\end{eqnarray}

Notice that $[x_i,x_j] = 2\pi_i\theta_i$; and in this sense one sometimes says
that the coordinates of the manifold do not commute.  Of course there is no
structure of product which is available for the coordinates, what the sentence
means is that, in fact, the space is not characterised any more by an algebra
of functions and instead requires an algebra of operators: out of such an
algebra one may extract a convenient ``basis'' which can be regarded as
noncommuting coordinates over an auxiliary object which keeps memory of the
product structure carried by the algebra.

Let us try to relate the above algebraic description with the quotient space
viewpoint.  We can for simplicity think of a $T^2$.  So the idea is to have a
foliation with angular coefficient $\theta$ and identify the points which lie
on the same leaf.
%space for figure

As already remarked last lecture this operation is quite tricky for $\theta
\not\in \Qbar$ and for example will force us to review our concept of
measurable functions (and consequently many tools we use all the time,
including Lebesgue integration) but we will not discuss these complicated
problems since anyway we will stick to a trivial geometrical setting, that is,
$\theta \in \Qbar$.

So let us consider the forms
\[T^2 = \Rbar^2/\Zbar^2\]
Let us consider the line bundle in $\Rbar^2$
\[y = \theta x + b\]
$\theta$ fixed, $b$ identifying a leaf in a non univocal way

Identify the points which lie on the same leaf and then quotient by $\Zbar^2$
thus wrapping the torus.

When two values of $b,b'$ identify the same leaf~?  When there exist $n \in
\Zbar$ such that
\[b' = b + \theta n\]
The quotient space viewpoint tells us: if you wish to see a $C^*$-algebra that
is rich enough to characterise the quotient space, you must not take the
structures of the parent space and have them inherited; instead take the graph of
the equivalent relation, ${\cal R} \subset {\cal K} \times {\cal K}$ (${\cal
K}$ = parent space), ${\cal R} \equiv \{(Z,Z'),Z,Z' \in {\cal K}, Z \sim Z'\}$
and work on that.

An element of the $C^*$-algebra is going to be a matrix
\[(a)_{Z,Z'}\ \ \mbox{indexed by points $(Z,Z') \in {\cal R}$}\]
The product of the algebra will of course be the matrix product
\[(ab)_{ZZ'} = \sum_{{Z''} \atop {Z \sim Z' \sim Z''}} (a)_{ZZ''}(b)_{Z''Z'}\]
Notice that such matrices may be infinite but will any way be with discrete
entries, because we are arranging (for various purposed beyond this
discussion) that the set of representatives for a given equivalence class is
denumerable.

And this is why we define functions over the leaf $F(b,b')$ or, if we prefer,
$F(b,n)$ where it is implied $b' \equiv b + \theta n$.

The matrix product will give rise to rewriting
\[F*G = \sum_n f(b,n) G(b + \theta n,m-n)\]
which is the $*$ operation first studied by CDS with the noncommutative
algebra already written.

\section{M theory on the rationally foliated torus}  

Now let us move to a $M$ theory setting.  We will build an example which
embodies our equivalence relation: here is how.

II A theory on a torus; compactification is along space like circle of
vanishing radius (Seiberg).\newline CDS considers a background 2-form
potential $\theta$.\newline We will study a similar setting with a square
torus; however we will have a $D0$-brane embedded in the torus and do a
$T$-duality.
\[\left.\begin{array}{ll}
\tau = i\\
P = \theta + ia\ \mbox{($a$ = area)}\end{array}\right) \mbox{original torus}\
\stackrel{T}{\Longleftrightarrow} \begin{array}{ll}P' = i\\
\tau' = \theta + ia\end{array}\ \mbox{(that is $a' = 1,\ \theta ' = 0$)}\]
%space for figure
Dual torus with the $D1$ brane embedded.

We may have fundamental strings as well.
%space for figure

And notice that if the length is minimized this induces an equivalence
relation $b' \sim b + n\theta$.

Consider a field operator which creates or annihilates such strings; its
arguments are two points which are equivalent in the sense of the foliation

{\it Natural to interpret such operators as belonging to the noncommutative
algebra of the foliated torus.}

Let us start with $\theta = 1/N$
%space for figure

For the dual torus\\
$\sigma \in [0,\Sigma]$

\begin{eqnarray*}
&&\tau' = \theta + i\Sigma h\\ &&p' = i
\end{eqnarray*}
The fundamental strings just described dynamically connect points $\sigma$ and
$\sigma + W(\Sigma/N)$ $(W \in \Zbar)$.  (From now on the vertical direction
will play very little role and we will just forget it).\newline We can thus
define nonlocal fields $\phi = \phi(\sigma,W)$ or if we prefer $\phi
= \phi(K,-)$ by means of a Fourier transform), $K,W
\in \Zbar$.  The energy of the corresponding modes is
\[E(K,W) = \left(\frac{1}{\Sigma^2} [K^2 + {\cal
W}^2]\right)^{1/2}\] Now notice that the very definition of the dual torus
$(T^2)$ allows us to identify, along the $\sigma$ direction which is the
interesting one (that is, on $T^1 = S$) the points $\sigma$ and $\sigma +
\Sigma n$, $n \in \Zbar$.  On the other hand we have just seen the
dynamical connection induced by fundamental strings between points $\sigma$
and $\sigma + W$ can suggest us to think of an auxiliary torus $T^1$, of
length $\Sigma/N$.
%space for figure

\parbox{4cm}{original torus} \hspace*{1cm} \parbox{4cm}{These connections arise from the
dynamic, that is, the background $1/N$} \hspace*{1cm} \parbox{4cm}{$N$ copies
of a new torus whose coordinate $x$ takes value $x \in [0,\Sigma/N]$}

Each of the ``arches'' is folded and becomes a smaller torus of its own.

One further comment can be made with relation to a choice of generic $\theta
\in \Qbar$, that is $\theta = P/N$, $P,N \in \Zbar$ and relatively
prime.  Then the equivalent relation coming from the background just induces a
relabeling of the order in which the smaller tori are attached: the sequence
0th, 1st, 2nd\ldots $N$th is replaced by 0th, $((P)_{\mod N})$th, $((2P)_{\mod
N})$th,\ldots$((P(N-1))_{\mod N})$th.

Passing from the original $T^1$ to the smaller torus has the effect of
changing the notion of integer and fractional momentum/winding quantum numbers.  In
particular we can rewrite
\begin{eqnarray*}
K &\equiv& KN + q\hspace*{3cm}q = 0,1,\ldots N-1\\ W &=& WN + (a -
b)\hspace*{1.5cm} a,b = 0,1,\ldots N-1
\end{eqnarray*}
($a-b$: this index says in which of the $N$ arches the string begins, resp.\
ends)

At the same time, the fields $\phi(\sigma,W)$ can be rephrased in terms of the
new variables, but we need to introduce indices to keep memory of which arch
we deal with:
\[\phi(\sigma,W) \longrightarrow \phi_{ab}(x,W)\hspace*{2cm} a,b = 1,\ldots
N\] We can rewrite that as $\phi_{ab}(x,y)$ if we interpret the winding mode
as $K-K$ momentum in an additional direction.

Energy of course becomes
\[E^2 = \frac{1}{\varepsilon^2} \left((NK + q)^2 + (NW + (a-b))^2\right)\]
Now our aim is to reproduce such spectrum, using the fields $\phi_{ab}$, by
putting appropriate boundary conditions over the fields $\phi$ themselves. For
our convenience let's present an explicit representation of the algebra of
operators on the $NC$ torus in the case where $\theta = 1/N$.
\[U = \left(\begin{array}{ccccccc}
0 & & 1 & 0 & 0 & 0 & \ldots\\ & & 0 & 1 & 0 & \ldots\\ & & & 0 & 1 & \ldots\\
& \phi & & & 0 & 1 & \mbox{}_{\ddots}\\ & & & & & 0 & \mbox{}_{\ddots}\\ 1 & &
& & & &
\end{array}\right)\ \ \ \ \mbox{shift matrix}\]

\[V = \left(\begin{array}{ccccc}
1 & & & \phi\\ & e^{2i\pi/N} & & \\ & & e^{4i\pi/N}& & \\ & \phi & &
e^{6i\pi/N} & \mbox{}_{\ddots}\end{array}\right)\ \ \ \ \mbox{phases matrix}\]

What happens after a turn to the fields $\phi$~?
\begin{eqnarray*}
&&\phi_{ab} \left(x + \frac{\Sigma}{N},y\right) = \phi_{a+1,b+1}(x,y)\\*[5mm]
&&\phi_{ab} \left(x,y + \frac{\Sigma}{N}\right) = \phi_{ab}(x,y)
\end{eqnarray*}
In matrix notation
\begin{eqnarray*}
&&\phi \left(x + \frac{\Sigma}{N},y\right) = U^+ \phi(x,y)\ U\\*[5mm] &&\phi
\left(x,y + \frac{\Sigma}{N}\right) = \phi(x,y)
\end{eqnarray*}
This is not the end of the story, the energy contains fractional contribution
for the momentum (they of course used to be integer on the big torus).

To mimick these, we introduce Wilson loops
\[W(x) = \exp\left(i \oint A_y(x,y)dy\right)\]
with $A_x = 0$ and $A_y$ independent of $y$:
\[W(x) = \exp\left(i\frac{\Sigma}{N} A_y(x)\right)\]
The background vector potential we choose is
\[W(x) = \exp\left(i \frac{x}{\Sigma}\right)\ \ \left(\begin{array}{cccccc}
\mbox{}^{\ddots} &  e^{-2\pi i/N} & & & \phi\\ &  & 1 & e^{2\pi i/N} & & \\
& \phi & & & e^{4\pi i/N} & \mbox{}_{\ddots}\end{array}\right)\] Notice that
\[W\left(x + \frac{\Sigma}{N}\right) = U^+ W(x) U\]
that is, the behaviour of the background for a complete turn on the small
circle is identical to the one of $\phi$.

Explicitating the vector potential
\[A_y = \underbrace{\frac{x}{\Sigma^2} N\ Id}_{\mbox{ABELIAN PART}} + \frac{1}{\Sigma}\ \
\left(\begin{array}{cccccc}
\mbox{}^{\ddots} & & & \phi & & \\
& -2 & & & &\\ &  & -1 & & &  \\ & & & 0 & & \\ &
\phi & & & 1 & \mbox{}_{\ddots}\end{array}\right)\]
\hspace*{5cm} $\Downarrow$\\
A unit of abelian magnetic flux through the torus.

What would happen if $\theta = p/N$~?
\[W(x) = \exp\left(ip \frac{x}{\Sigma}\right)\ \ \left(\begin{array}{cccccc}
\mbox{}^{\ddots} & e^{-2\pi i/N} & & & \phi & \\
& & 1 & & & \\ & & & e^{2\pi i/N} & & \\ & \phi & & & e^{4\pi i/N} &
\mbox{}_{\ddots}\end{array}\right)\]
The abelian part of the magnetic flux is multiplied by $p$ and so happens to
the magnetic flux through the torus.

Now let's see what happens to the action when we pass from the variables on
the original torus to the ones on the small torus.  The part of the lagrangian
which may encounter problems is the free one, because we did an operation of
``cut'' over the torus.  The free dynamical evolution ``does not recognize''
that we have reasons to identify points of a distance of, let's say,
$\Sigma/N$; she just states that after the first arch there comes the second
one (not that the first arch will close on itself to form a torus~!).  On the
other hand the interaction terms will understand that fact very well, since it
is the interaction which gave us reasons for the whole operation\ldots  In
other terms, the polinomial terms in the action, as soon as they are
appropriately put on the deformed $NC$ torus, will have to be correctly
translated in terms of the deformed product $*$.  so we actually don't even
need to check the interaction terms.  As for the free term it is
understandable that the reformulation on the small torus, which involves a
``cut and glue'' procedure, will create fictitious magnetic monopoles, whose
role is exactly to mimick the ``jump'' thus induced.

Anyway let's see directly what happens.
\[{\cal L}_{\mbox{\footnotesize free}} = \int dx\,dy\,Tr\left(\dot\phi^2 -
(D\phi)^2\right)\hspace*{2cm} \ddot\phi = D^2\phi\] with $D\phi = \partial
\phi + i[A,\phi]$.
\[D^2 = D_x^2 + D_y^2 = (\partial_x)^2 + \left(\partial_y +
i[A_y,\cdot]\right)^2\] $\partial_x \rightarrow$ this part of the spectrum
should be evaluated on the small torus of length $\Sigma/N$.\newline  But we
imposed conditions on $\phi$ such that it comes back to its original value
only after $N$ turns, so it's like the torus had size $\Sigma$.\newline
(Obvious, the $x$ direction is the original one\ldots)\newline Spectrum
contribution: $i/\Sigma[KN + p]$ ($p = 1,\ldots N-1$)\newline $\partial_y
\rightarrow$ $i(NW/\Sigma)$\newline
$A_y,\cdot \rightarrow$ Let's see what the commutator is like
\[A_y,\phi]_{a,b} = \frac{1}{\Sigma} (a - b)\phi_{ab}\]
That is
\[E^2 = \left[\left(\frac{kN + p}{\Sigma}\right)^2 + \left(\frac{NW}{\Sigma}
+ \frac{a - b}{\Sigma}\right)^2\right]\] as required.

To summarize, we have now a prescription for rewriting a theory over a
noncom.\ torus of characteristic parameter $\theta$ into another $SYM$ theory
with appropriate size of the gauge group and number of magnetic changes.

\begin{center}
Abelian $SYM$ theory over a $NC$ torus of size $\Sigma$ with parameter $\theta
\simeq p/N$\\
$\Updownarrow$\\ $U(N)\ SYM$ on a torus of size $\Sigma/N$ with $p$ units of
magnetic flux.
\end{center}
The $SYM$ content of this statement is non trivial and allows us, for example,
to relate theories with very different values of $N$ (interesting, especially
in view of large $N$ limit conjectures).

The $NC$ content is instead quite trivial.  It leaves us with the opinion that
switching on the noncommutativity corresponds to introducing a few magnetic
monopoles (and of course we knew about that).

In the next lecture we will see another situation in which a $NCG$ system can
be again completely described in terms of a magnetic field mechanism.  So we
will be aware that in many cases of interest in brane physics the $NC$
deformation does not go much beyond the good old $B$ field.  Of course we do
not believe this to be the end of the story and we expect to find many
examples of different characteristics, we will comment further on that.
\section*{Lectures 3-4}
The noncommutative torus in string theory: how to relate noncommutativity 
and magnetic field background. A situation in which ``deforming the 
geometry'' barely means ``switching on a $B$ field''. 

Mimicking a ``noncommutative geometry'' by appropriate $B$ fields. 
Regularization properties of deformation into noncommutative structure 
of the Feynman diagrams. 

\section*{Introduction} 
We will now proceed to discuss another example in which the noncommutativity
again appears as a rephrasing of well known physics, namely the physics of the
magnetic field, by means of a convenient mathematical formulation.  The
reference is a joint work D.B.\ and L.~Susskind hep-th/9908056.

Our aim will be to study the perturbative structure of noncommutative YM
theories.  We will then discuss the vertex structure for light cone string
theory in presence of a $D3$-brane and a background $B$ field; we will find
that it has Moyal phrases typical of NCG systems.  We will moreover show how
such a vertex structure is mimicked by a model of a pair of opposite electric
charges in a magnetic field.

As in the previous lecture, we will find interesting results from the
viewpoint of string theory, and in particular about the regularization
properties of the noncommutative deformation over the Feynman diagrams. Namely
we will be able to prove that for planar diagrams the Feynman amplitudes are
the same as for the corresponding commutative YM theory; instead generic non
planar diagrams are regulated by switching on the noncommutativity.  We can
also argue that the large $N$ limit is the same for both ordinary and $NC$
theory, accounting for thermodynamical properties of the theory not to be
changed, and this is a result of some interest for string theory (and in
particular in connection with the proposed AdS/CPT correspondence).

What should be stressed here is that we are confronted once more with a result
which is negative from the point of view of noncommutative geometry: we have
another example in which the deformation induced by noncommutativity (at least
most NCG can be thought of as deformation of ordinary geometrical spaces; not
the space of inequivalent Penrose tilings, but anyway a bunch of interesting
examples) can be rephrased as some variant of background $B$ field.  Our
spirit is of course to explore how far we can go along this vocabulary with
the ultimate intention of builiding examples of NCG which are not reducible to
$B$ fields and hopefully will talk about some new and interesting ``physical''
phenomenon, we will go back to this point.

\section{String theory setup} 

Let's start with our setup, this time we have bosonic string theory, we have a
brane $D3$ oriented along $X^0$, $X^1$, $X^2$, $X^3$ and also we have a
background $B^{\mu\nu}$ along direction 1,2 in light cone frame.
\[X^\pm = X^0 \pm X^3\hspace*{2cm} X^* \equiv \tau\]
\[{\cal L} = \frac{1}{2} \int_{-L}^L d\tau\,d\sigma
\left[\left(\frac{\partial x^i}{\partial\tau}\right)^2 - \left(\frac{\partial
x^i}{\partial\sigma}\right)^2 + \underbrace{B\, \epsilon_{ij}}_{B_{ij}}
\left(\frac{\partial x^i}{\partial\tau}\right) \left(\frac{\partial
x^j}{\partial\sigma}\right)\right]\] This is the lagrangian for the open
string sector (of strings with extremal points as the $D3$ brane).

Do rescaling:
\begin{eqnarray*}
x^i &=& y^i/\sqrt B\\
\tau &=& t\,B
\end{eqnarray*}
\[{\cal L} = \frac{1}{2} B \int_{-L}^L dt\,dB \left[\left(\frac{1}{B^2}
\left(\frac{\partial y^i}{\partial t}\right)^2 - \left(\frac{\partial
y^i}{\partial\sigma}\right)^2 + \epsilon_{ij} \left(\frac{\partial
y^i}{\partial t}\right) \left(\frac{\partial
y^j}{\partial\sigma}\right)\right)\right]\] This is a more convenient form
because we are interested in a regime with very large $B$, so we will drop the
first term.

Now we manipulate the lagrangian and solve the classical eq.\ of motion, this
gives
\[y(\sigma) = y + \frac{\Delta\sigma}{L}\]
That is
\[\mbox{\fbox{${\displaystyle {\cal L} = \frac{1}{2}
\left[-2\frac{\Delta^2}{L} + \dot y^i \epsilon_{ij} \Delta^j\right]}$}}\]
$-2\Delta^2/L$ = ``harmonic oscillator term''\\ $\dot y^i \epsilon_{ij}
\Delta^j$ = ``Landau level term''

What original form for a lagrangian.  Shall we see if we can mimick that by
means of a system containing harmonic oscillators and charges inside a strong
magnetic field~?

We are just going to see that this is, indeed, possible.

\section{A mechanical model}
The model consists of a spring with elastic constant $K$
at whose ends we have two particles of mass $m$ and electric charge $e$, 
embedded in a magnetic field.  

$|\vec B|$ high; $|\vec B|$, $e$, $m$ such that Coulomb and radiation terms
are negligible.  The system is essentially confined to a plane.
\[{\cal L} = \frac{m}{2} \left((\dot x_1)^2 + (\dot x_2)^2\right) +
\frac{Be}{2} \epsilon_{ij} \left(\dot x_1^i x_1^j - \dot x_2^i x_2^j\right) -
\frac{K}{2} (x_1 - x_2)^2\]
$((\dot x_1)^2 + (\dot x_2)^2)$: this term we neglect because we assume to
always be in the lowest Landau level.

Define C.M.\ and relative coordinates.
\begin{eqnarray*}
\vec X &=& (\vec X_1 + \vec X_2)/2\\
\vec \Delta &=& (\vec x_1 - \vec x_2)/2
\end{eqnarray*}
\[\longrightarrow {\cal L} = 2Be\,t_{ij} \dot X^i \Delta^j - 2K(\Delta)^2\]
Same form than the original lagrangian provided we choose appropriately the
parameters.

This is already interesting: the message is that a quantum of NCYM can be
described as an unexcited string stretched between two opposite charges inside
a strong magnetic field.

There are more remarkable things.
\begin{eqnarray*}
&&{\cal H} = 2K(\Delta)^2\ \ \mbox{(no contrib.\ as we know from magnetic
term)}\\*[5mm] &&p_i \equiv \frac{\partial{\cal L}}{\partial \dot X^i} =
2Be\,\epsilon_{ij} \Delta_j\\*[5mm] &&\rightarrow {\cal H} = 2K
\left(\frac{p^2}{2eB}\right)^2 = \left(\frac{K}{e^2B^2}\right)
\frac{p^2}{2}\\*[5mm]
&&\mbox{Looks like a nonrelativistic particle of mass}\ \ M = \frac{B^2e^2}{K}
\end{eqnarray*}
However the relation $p_i = 2B\,\epsilon_{ij} \Delta_j$ tells us that
$|\vec\Delta|$, the spatial extension of the system, is proportional to $|\vec
p|$, the momentum in the orthogonal direction. 
That is, you boost the system along a direction and it will spread in the
orthogonal one. This means that the particle is not pointlike any
more: how will the interactions reflect this~?

Suppose the charge 1 interacts locally with some potential for which the
charge 2 is transparent.
\[V(\vec x_1) = V(\vec X + \vec \Delta) = V\left(\vec X - \frac{1}{2B}
\epsilon^{ij} p_j\right)\]
Notice this shift of a momentum dependent amount.

Actually it's not obvious that such expression is well defined, $V$ is a
quantum operator and quantum ordering is important, since a priori the two
addends of the argument do not commute.  However it works.
\begin{eqnarray*}
&&V(x) = \int dq\,\tilde V(q) e^{iqx}\ \ \mbox{if $V$ admits Fourier transform
$\tilde V$}\\*[5mm] &&\Rightarrow V\left(x - \frac{\epsilon p}{2B}\right) =
\int dq\,\tilde V(q)\,e^{iq(x - \epsilon p/2B)}\\
&&\hspace*{5cm}\mbox{this is meaningful because $[q\cdot X,q \in p] = 0$}
\end{eqnarray*}
Now take $|K\rangle$ and $|\ell\rangle$ eigenstates for momentum:
\begin{eqnarray*}
&&\left\langle K\left| e^{iq(X - \epsilon p/2B)}\right|\ell\right\rangle =
\left\langle K\left|e^{iqx} e^{-iq(\epsilon
p/2B)}\right|\ell\right\rangle\\*[5mm] &&= e^{iq(\epsilon p/2B)} \left\langle
K\left|e^{iqx}\right|\ell\right\rangle = e^{-iq\,\epsilon p/2B} \delta (K - q
- \ell)
\end{eqnarray*}
This piece of the vertex is the novelty of noncommutativity.  And it is also
something familiar: these are the momentum-dependent Moyal phases.

That is, we have reproduced the Moyal phases characteristic of NCG by means of
this mechanical system which surely has no further content than the Landau
levels.  This is quite remarkable, the model didn't look particularly
illuminating or deep.

Next lecture we will work further on this example: we will study the
perturbative expansion of the theory in terms of Feynman diagrams and discuss
the regularization properties carried by the Moyal phases discussed above.  We
will find out as already anticipated that the planar diagrams (which we expect
to dominate in the large $N$ limit) are not changed while the non planar
diagrams are regularized (for a generic diagram).

\section{The interaction vertex}
%%%ZZZZZZZZZZZZZZZZZZZZZZZZ
As we have seen the basic quantum can be represented as a nonrelativistic 
object which is however not pointlike. So The basic interaction will be like a 
joining and splitting of strings, but always straight strings: 
%%% FIGURE 
We just saw that this interaction mimicks the Moyal phases of NCG. 
\par The propagator will be represented in double line notation: 
%%% FIGURE 
This should not be taken too literally as a double line like the quarks... 
This is not LQFT. It is always to be remembered that interactions are 
nonlocal (we are mimicking string theory which is of course nonlocal). 
\begin{eqnarray} \displaystyle{
Vertex \: = \: commutative \: vertex \:  e^{i p \theta  \varepsilon^{ij} q^j} 
} \end{eqnarray} 
\begin{eqnarray} \displaystyle{
e^{i p_1 \wedge p_2} = e^{i p_2 \wedge p_3} = e^{i p_3 \wedge p_1} 
} \end{eqnarray} 
where $p_a \wedge p_b$ indicates an antisymmetric product.   
\par Notice: 
\begin{itemize}
\item cyclic order matters. 
\item only cyclic order matters. 
\end{itemize}

Notice also channel exchange invariance. 
%%%%%%%%%FIGURE 

\begin{eqnarray} \displaystyle{
e^{i(p_1 \wedge p_2)} e^{i (p_1 +p_2) \wedge p_3} = e^{i (p_1
\wedge p_2 + p_2 \wedge p_3 + p_1 \wedge p_3)} } \end{eqnarray} 

\par Moreover any planar diagram can be rewritten as 
shown in the figure 
after appropriate sequence of duality moves. 
So it's enough to check Moyal phases contribution of the tadpole diagram 

\begin{eqnarray} \displaystyle{phase \: = \: e^{i k \wedge k} \: = \: 1}
\end{eqnarray}

{\bf No contribution of Moyal phases.}
\par In particular, structure of divergences is identical to the commutative 
case. 
\par Nonplanar diagram divergences are usually improved. An example is the 
diagram of the last figure. The commutative analogue has a logaritmic 
divergence. The improved diagram converges in this case. 
\par Typically, the diagrams undergo divergences improvement: this is indeed the case 
unless they contain a divergent planar subdiagram or unless the Moyal phases vanish 
because momentum in the 1,2 plane is vanishing. 

\section*{Conclusions}
\begin{itemize}
\item For planar diagrams NC YM theory has the same divergences as the corresponding
SYM.\newline In large $N$ limit NC YM = YM.\newline  In fact, it is claimed by
AdS/CFT arguments that introduction of noncommutativity does not charnge
thermodynamics of the theory in such correspondence.

(Actually here NCG is magnetic field~!  But this is anyway interesting -- it
was noticed before, that switching on $B$ does not change planar Feynman 
diagrams.  So at
least it is useful as a rewriting.)
\item Generic non planar diagrams are regulated.  That is, as high momenta it
looks that planar diagrams are in control of the overall behaviour.\newline
Beware: high momenta $\neq$ short distances~!!!  The system ``opens up'' in
the orthogonal direction when it acquires a momentum.
\item On a torus we should be much more cautious.\newline
Usually QFT on open space looks like QFT on a torus, provided we study modes
of high enough momentum.  Switching on NC crucially changes this.  For
example, if the background on the torus is rational, other processes are
allowed: endpoints of strings do not need to coincide, I have showed you
processes in which they are shifted, let's say, $n(\Sigma/N)$ if the size of
torus is $\Sigma$.  In fact, it was found by Krajewski and Wulkenhaar that non
planar diagrams diverge on a torus with rational background $\theta$.
\end{itemize}

To finish with, I propose a question: \\ 
How general is the description of a quantum of NC YM as a couple of opposite
charges elastically tied together in a strong magnetic field~?  Can we find
examples of NCG which describe a physics beyond the Landau levels~? \\ \\ \\ 

{\large New examples are needed. }

\section*{Acknowledgements}
It is a pleasure to thank the organizers of the TMR school who provided a 
wonderful opportunity to present and learn new subjects according to a 
format which is turning into a successful tradition. It is a pleasure as 
well to thank all the particpants who have brought in a good deal of 
interest and enthusiasm. 
\par The author is grateful to Anita Raets for typing part of the manuscript.

\newpage

\section*{Figures}

\begin{figure}[h]
\epsfxsize=150mm %75mm
\epsfbox{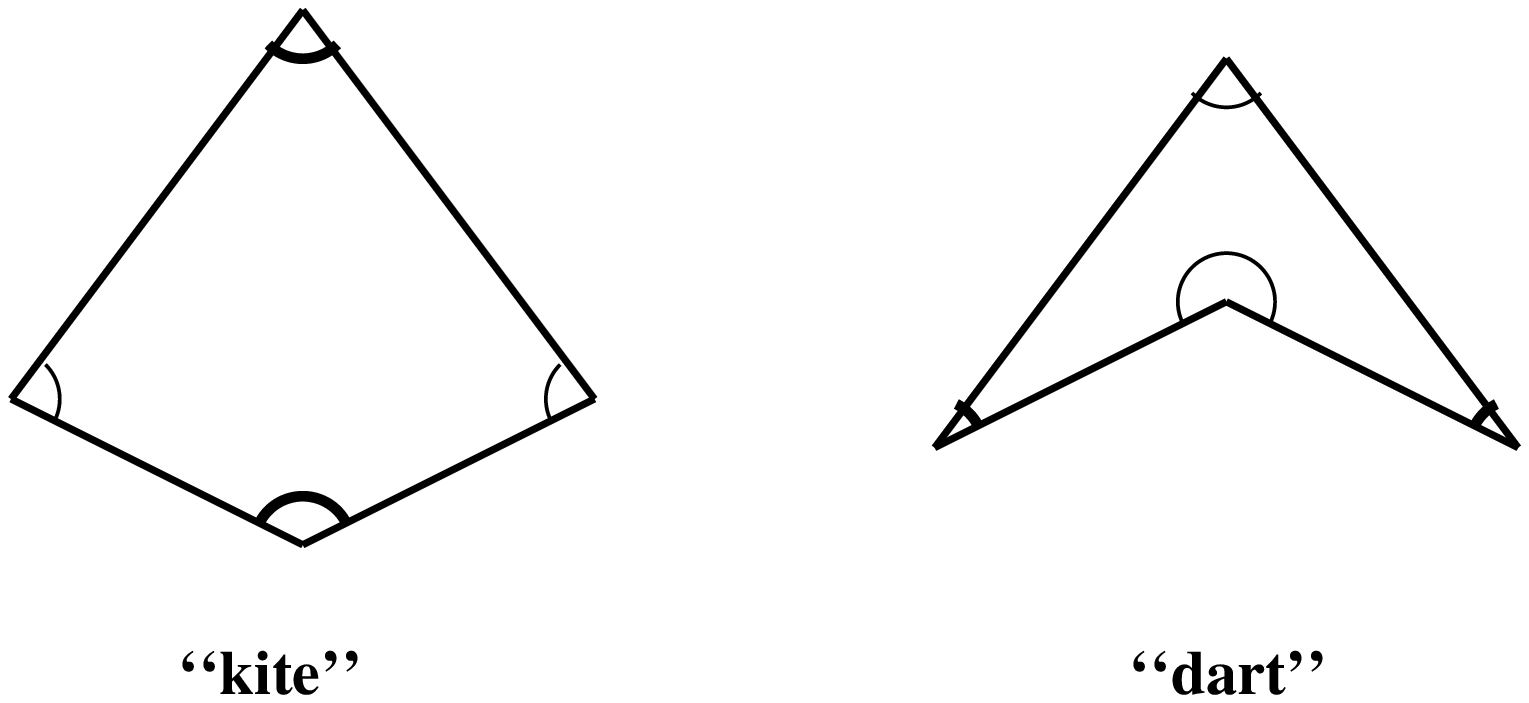}
\end{figure}
\indent {\large Figure 1.} \\

\newpage

\begin{figure}[h]
\epsfxsize=150mm
\epsfbox{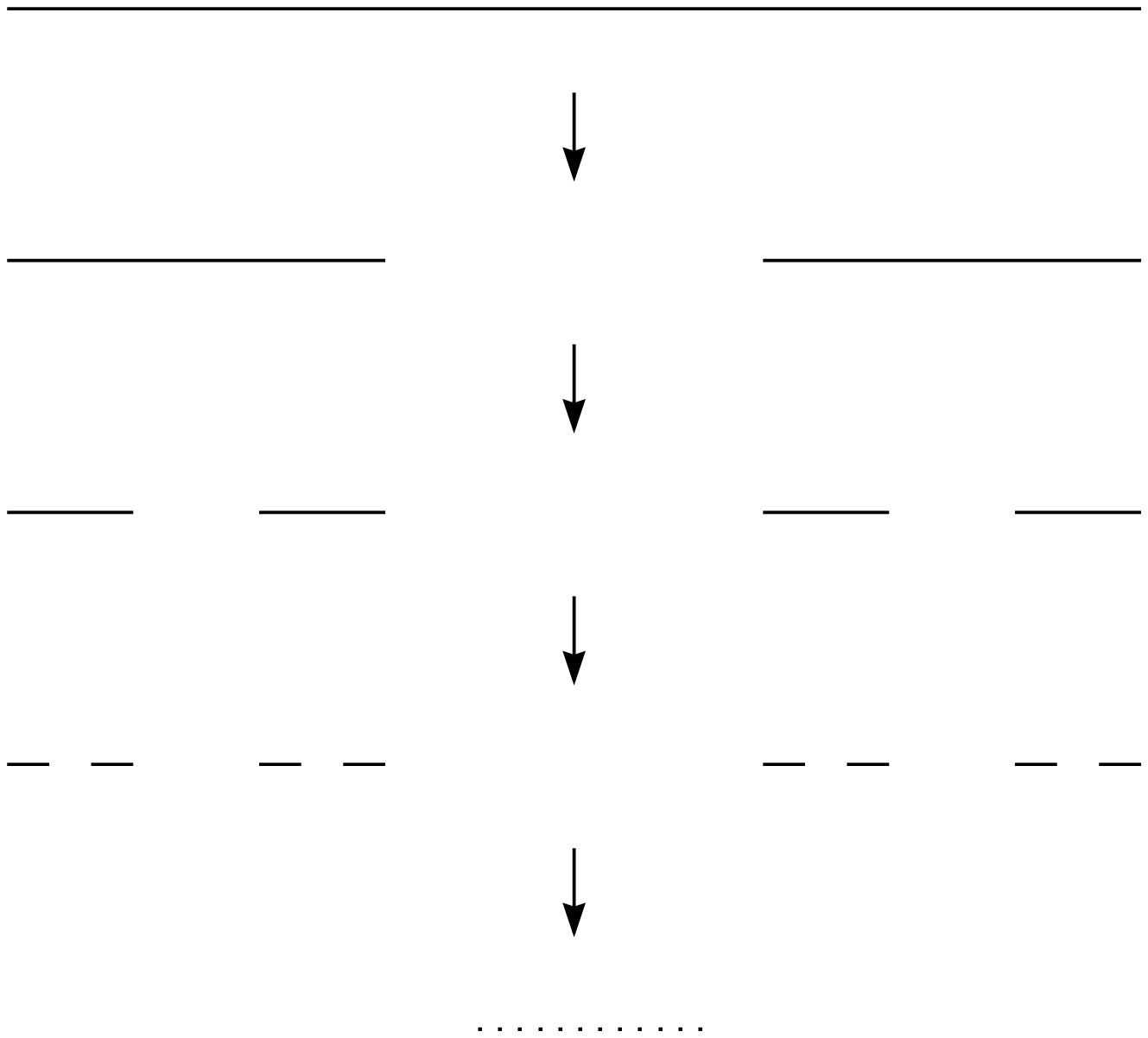}              
\end{figure}
\vspace*{15mm}
\indent {\large Figure 2.} \\

\newpage

\begin{figure}[h]
\epsfxsize=150mm
\epsfbox{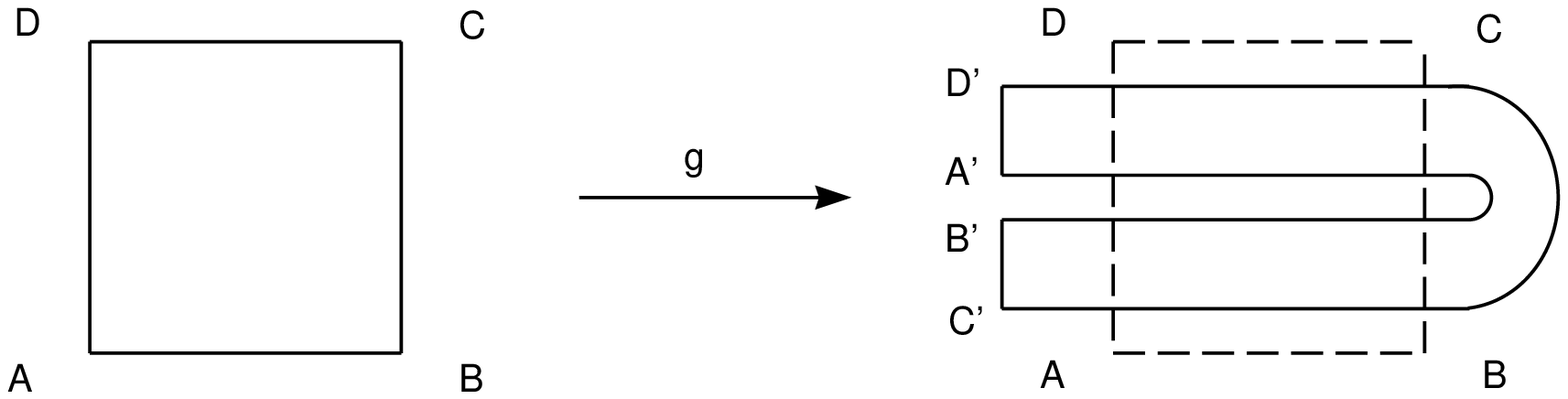}
\end{figure}
\indent {\large Figure 3.} \\

\vspace*{25mm}
%\newpage

\begin{figure}[h]
\epsfxsize=150mm
\epsfbox{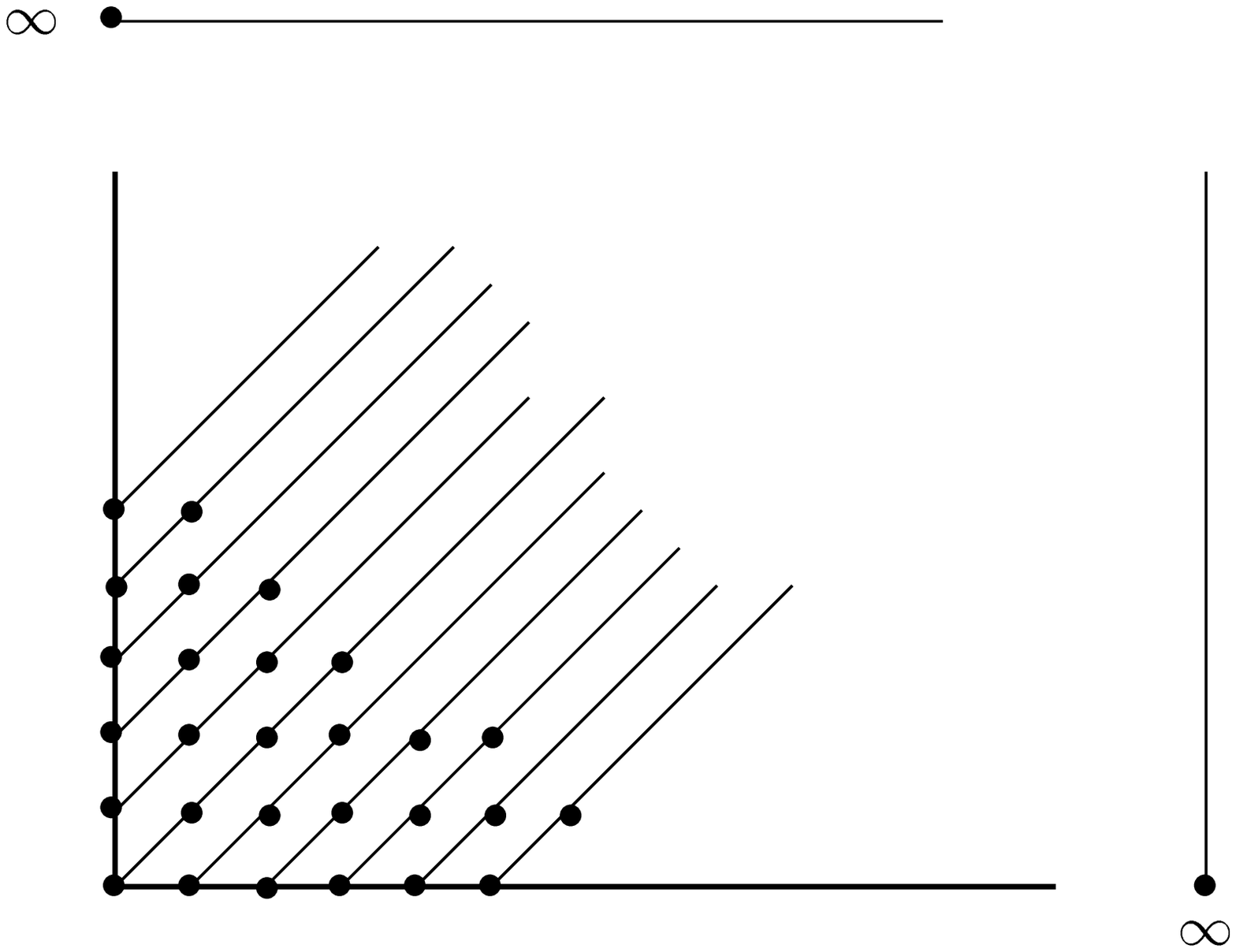}
\end{figure}
\indent {\large Figure 4.} \\

\begin{figure}[h]
\epsfxsize=150mm %75mm
\epsfbox{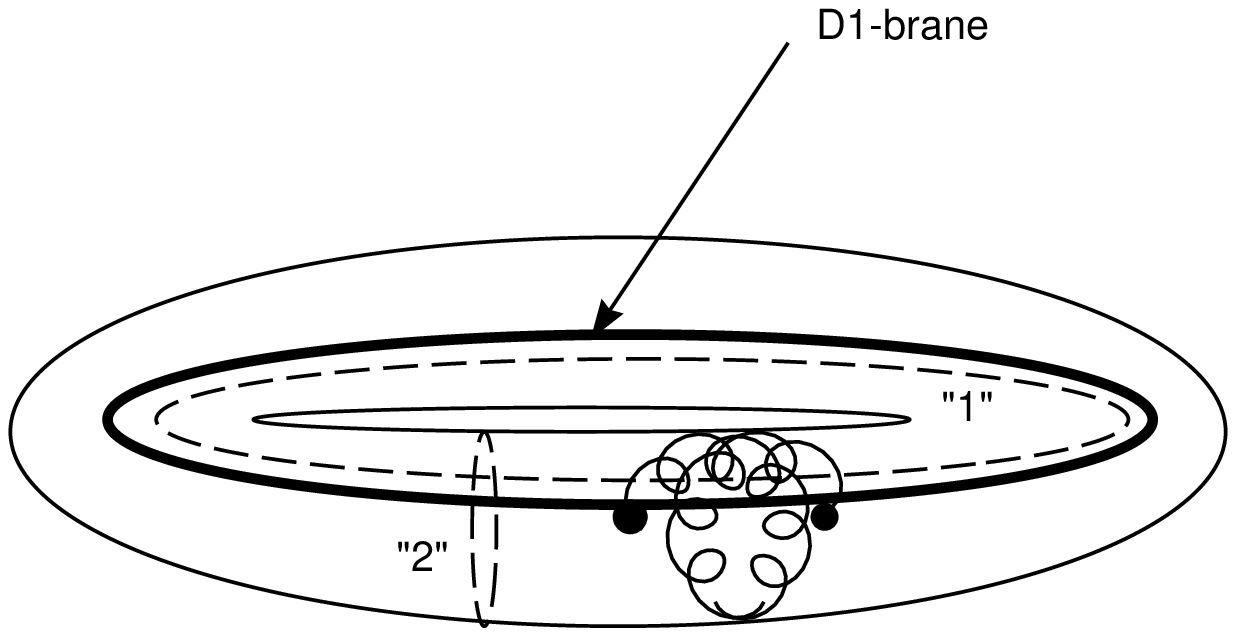}
\end{figure}
%\indent {\large Figure 1.} \\   

\begin{figure}[h]
\epsfxsize=150mm %75mm
\epsfbox{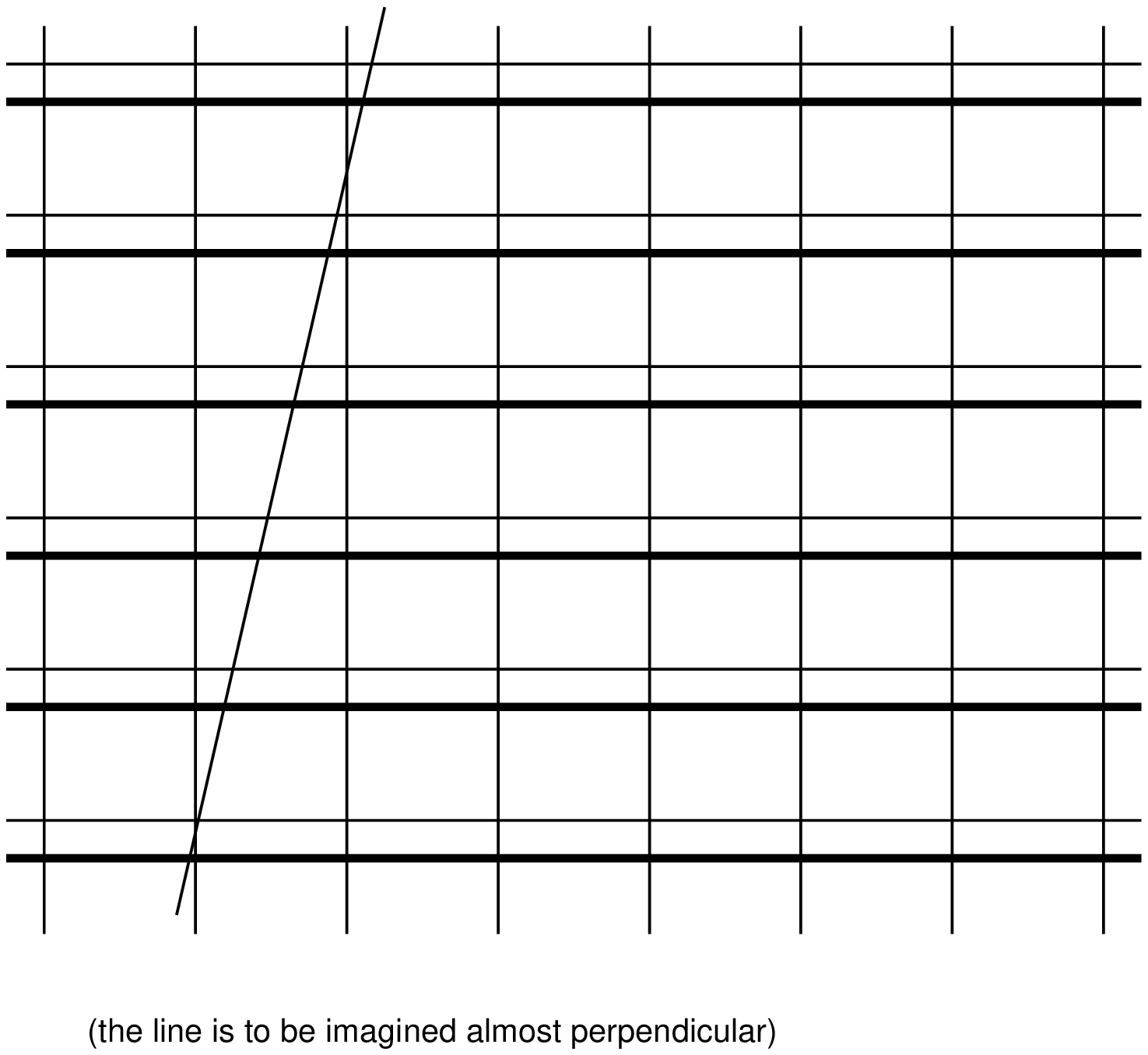}
\end{figure}
%\indent {\large Figure 2.} \\

\begin{figure}[h]
\epsfxsize=150mm %75mm
\epsfbox{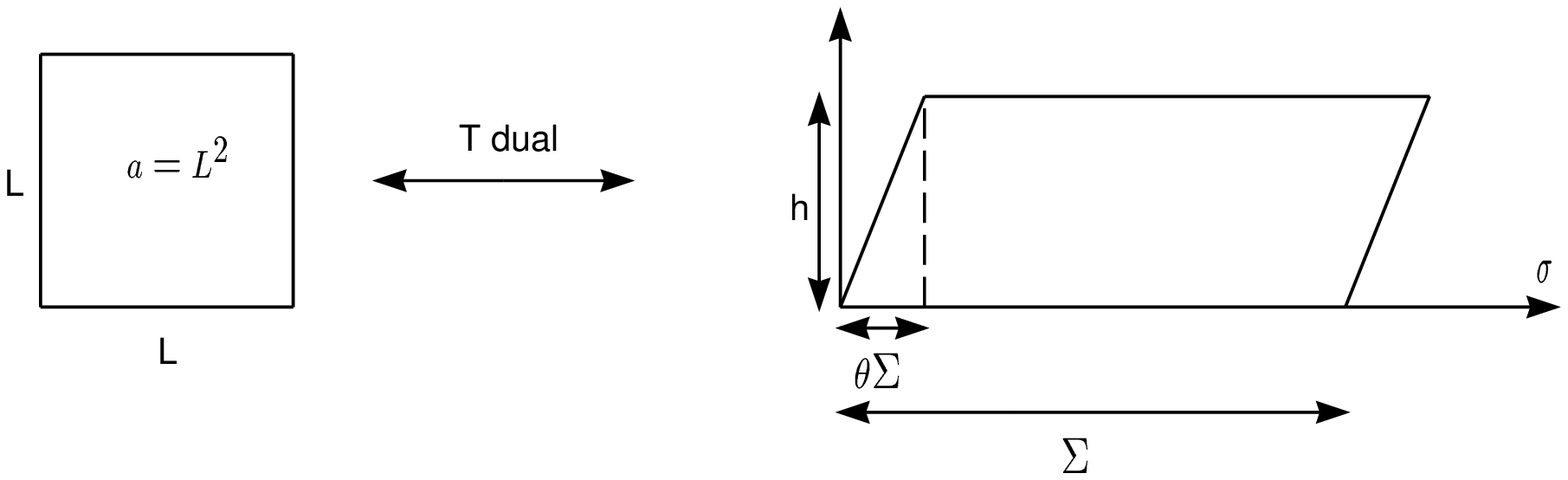}
\end{figure}
%\indent {\large Figure 3.} \\

\begin{figure}[h]
\epsfxsize=150mm %75mm
\epsfbox{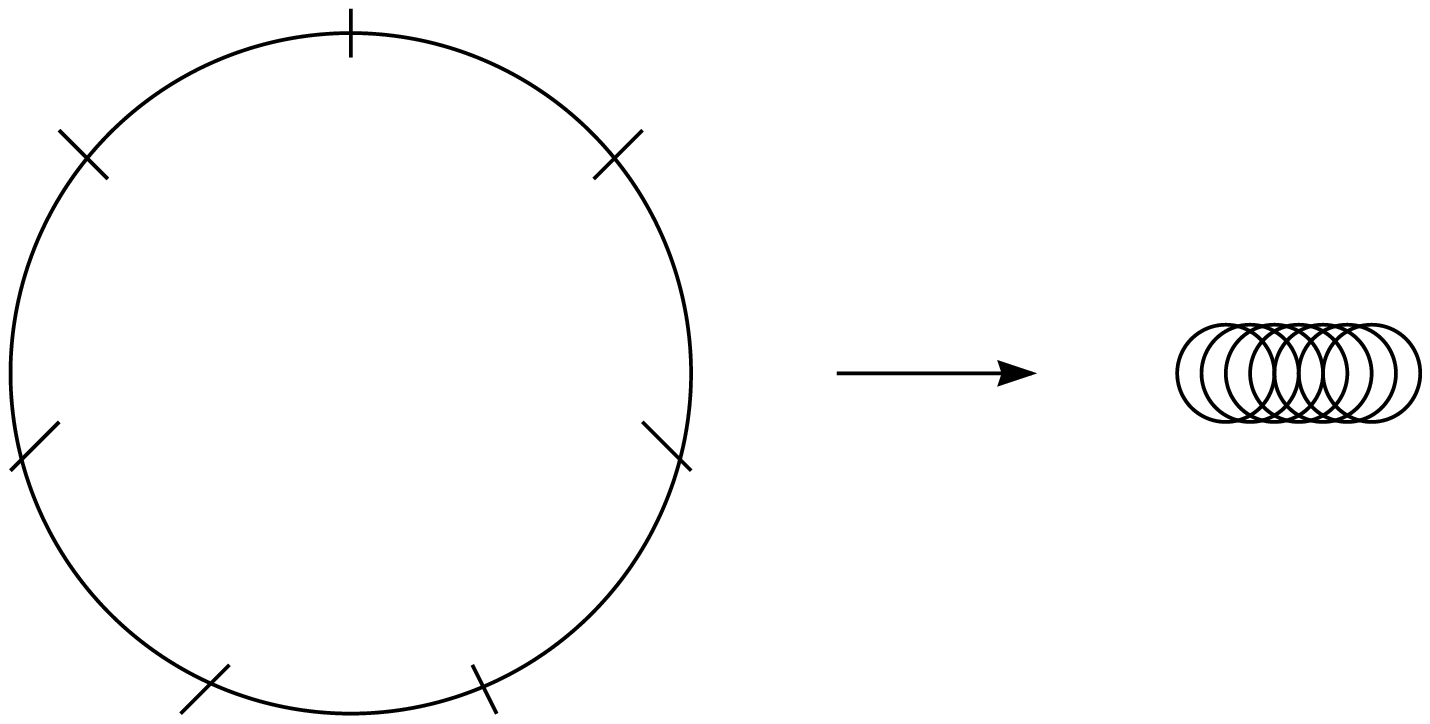}
\end{figure}
%\indent {\large Figure 4.} \\

\begin{figure}[h]
\epsfxsize=150mm %75mm
\epsfbox{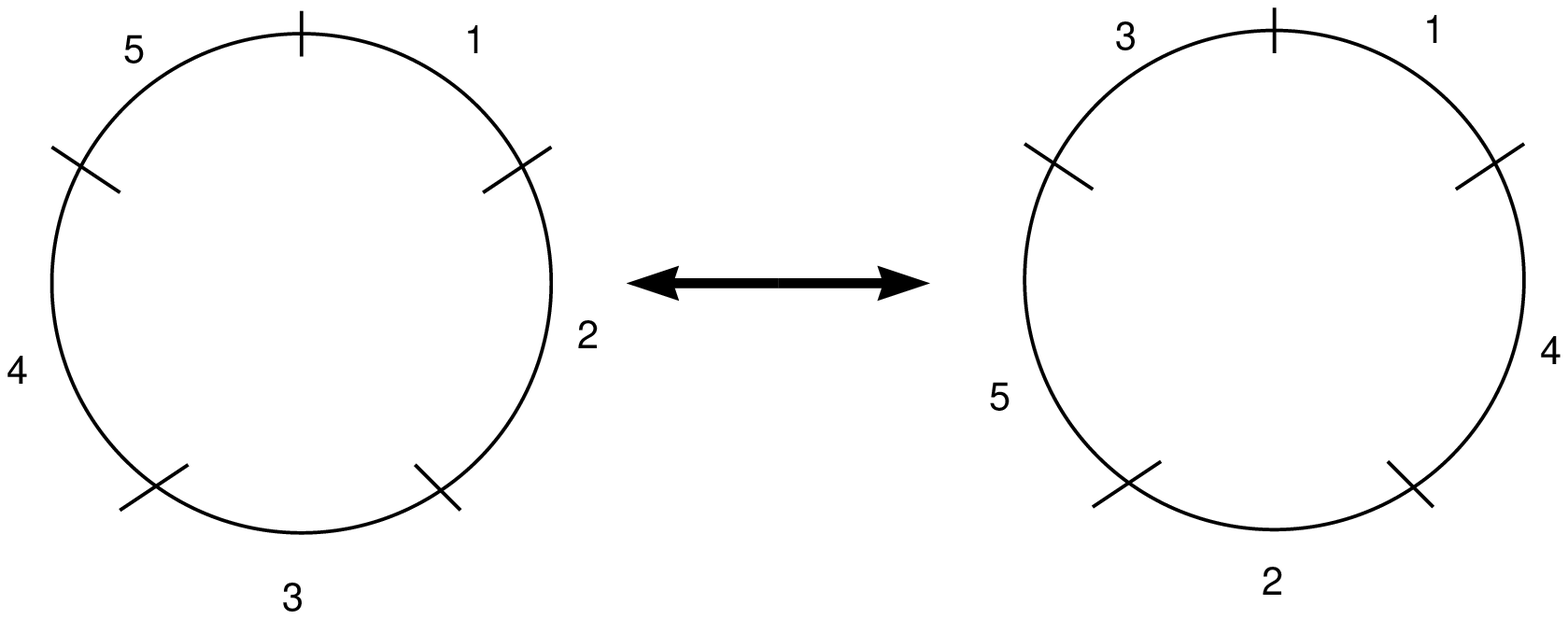}
\end{figure}
%\indent {\large Figure 5.} \\           

\begin{figure}[h]
\epsfxsize=150mm %75mm
\epsfbox{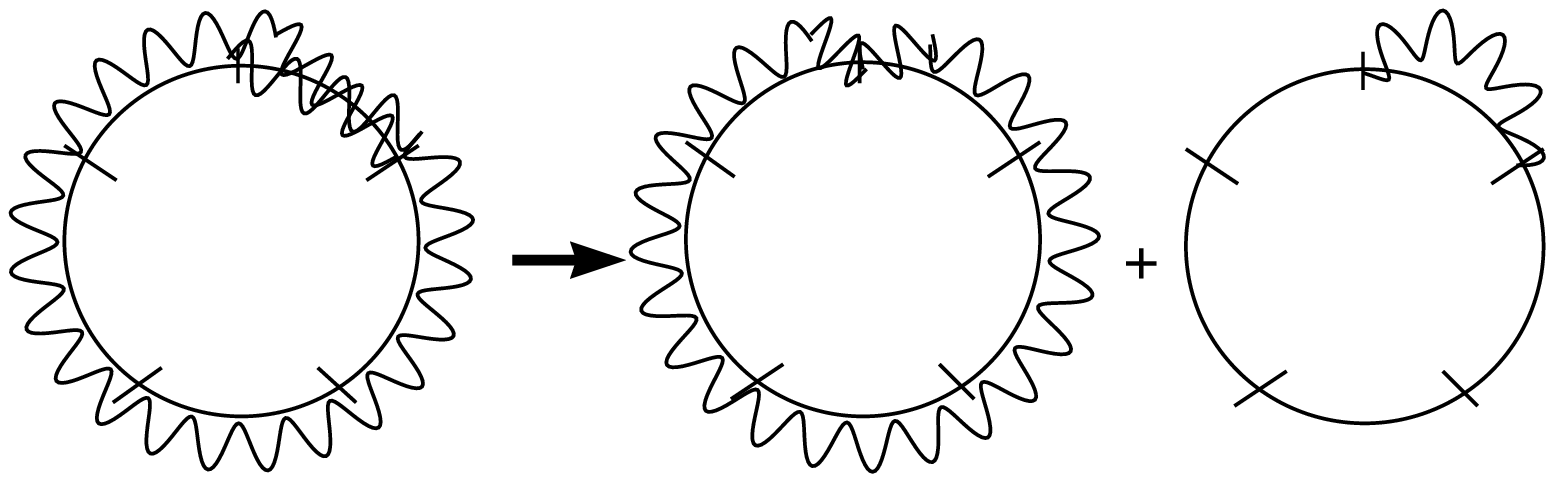}
\end{figure}
%\indent {\large Figure 6.} \\

\begin{figure}[h]
\epsfxsize=150mm %75mm
\epsfbox{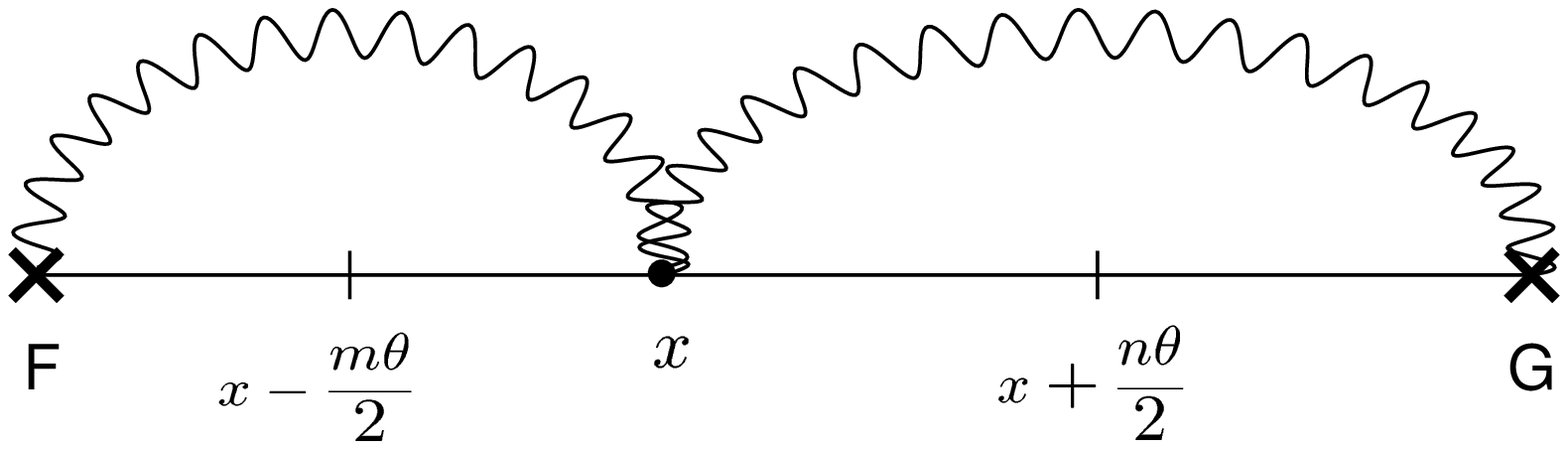}
\end{figure}
%\indent {\large Figure 7.} \\

\begin{figure}[h]
\epsfxsize=150mm
\epsfbox{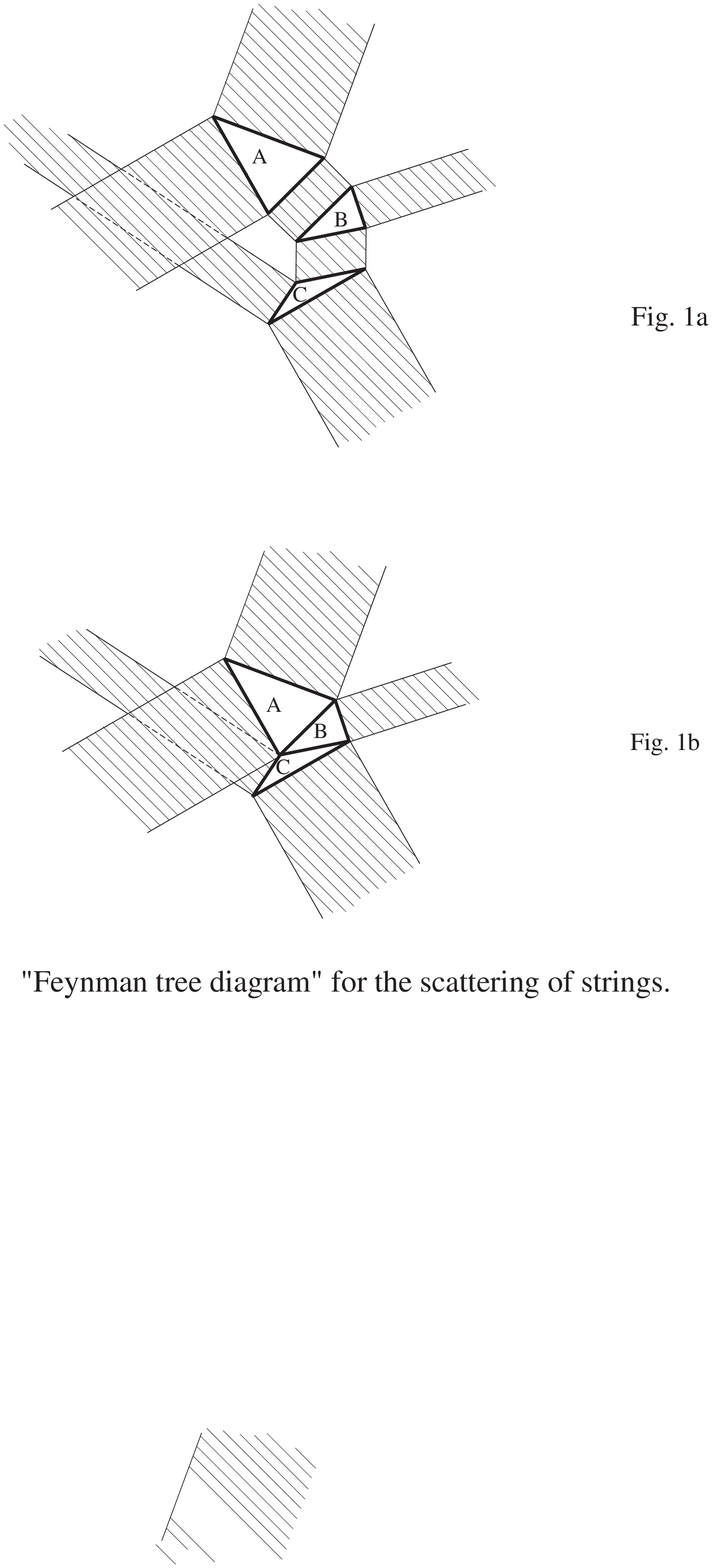}
\end{figure}    

\begin{figure}[h]
\epsfxsize=150mm
\epsfbox{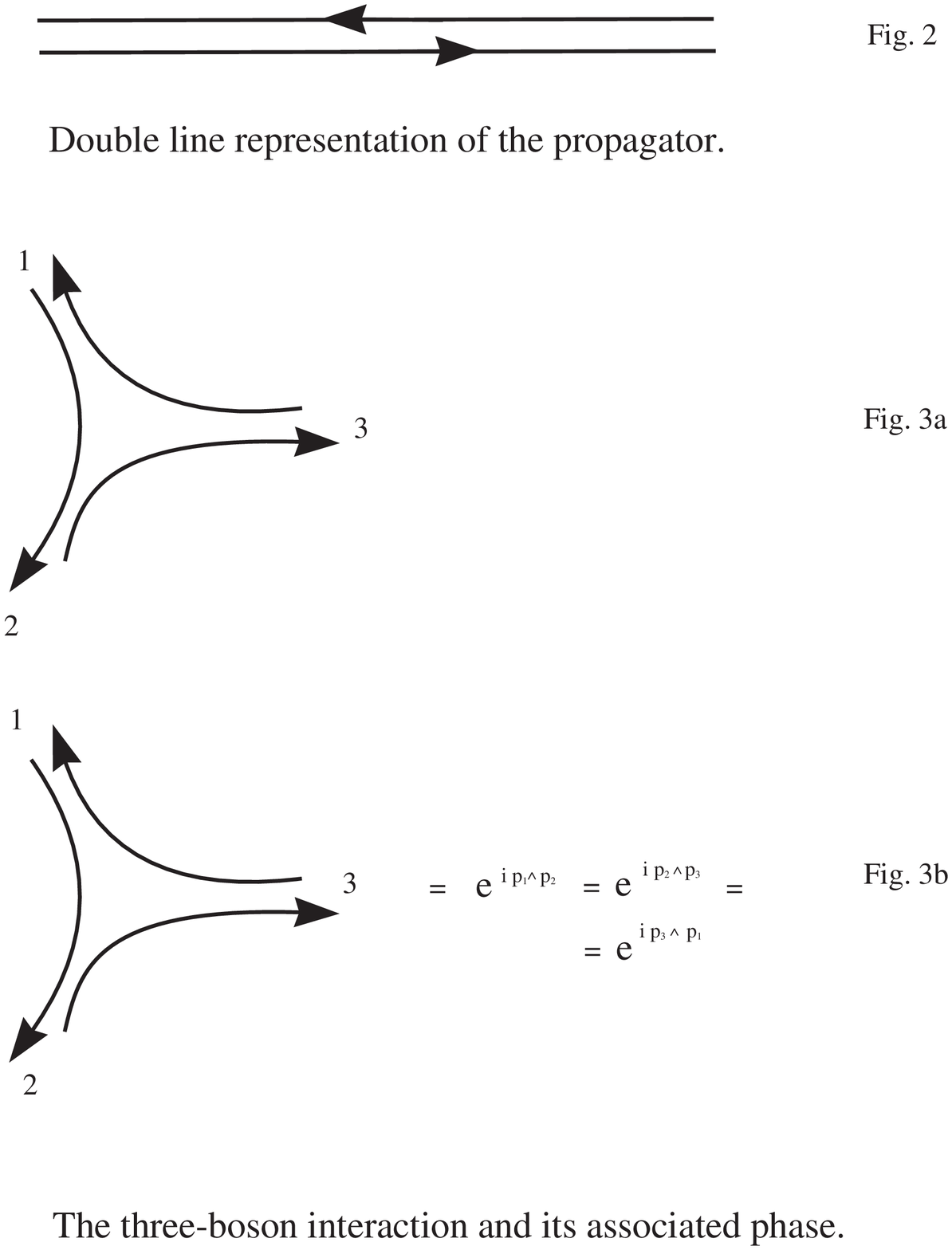}
\end{figure}

\begin{figure}[h]
\epsfxsize=150mm
\epsfbox{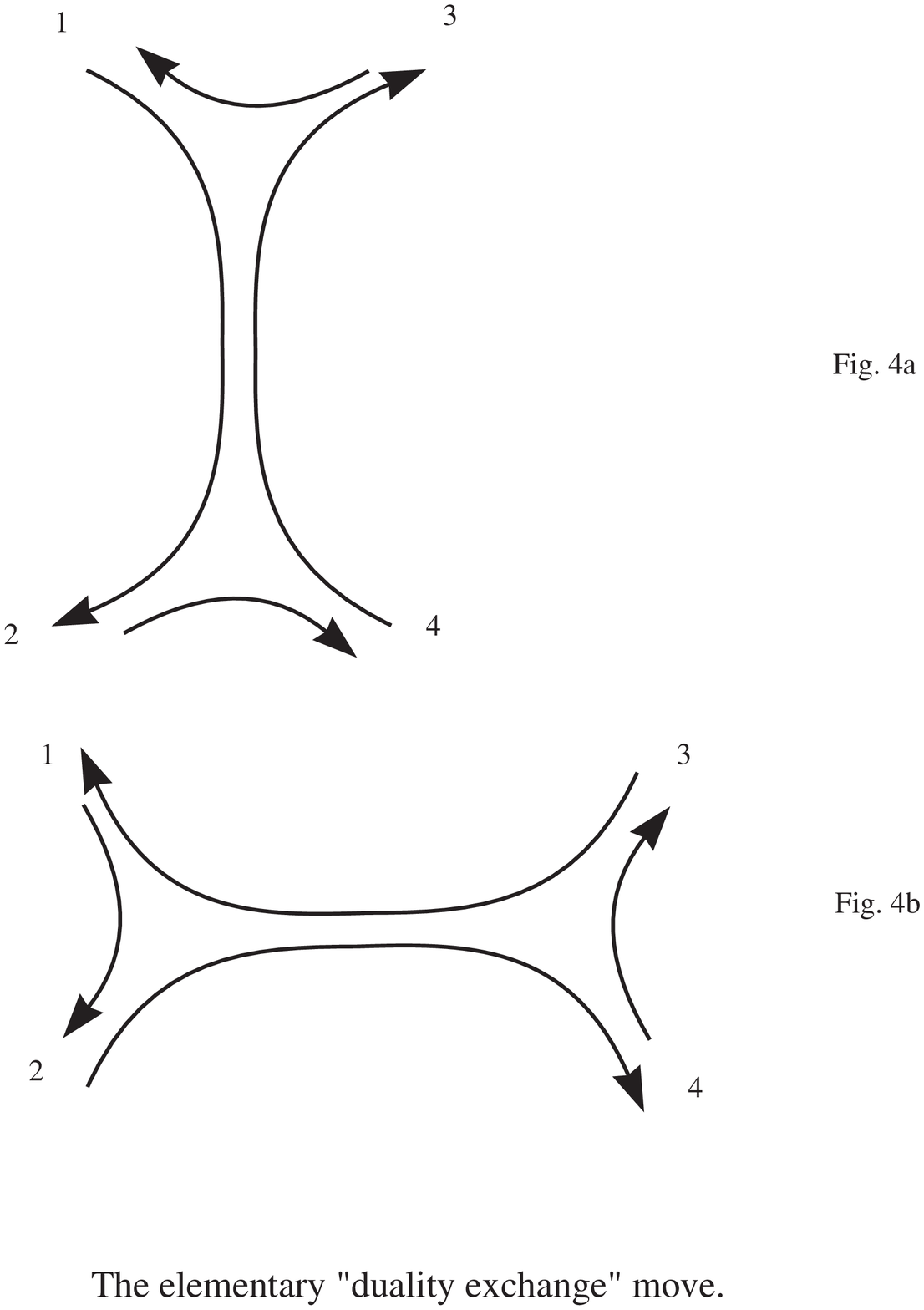}
\end{figure}

\begin{figure}[h]
\epsfxsize=150mm
\epsfbox{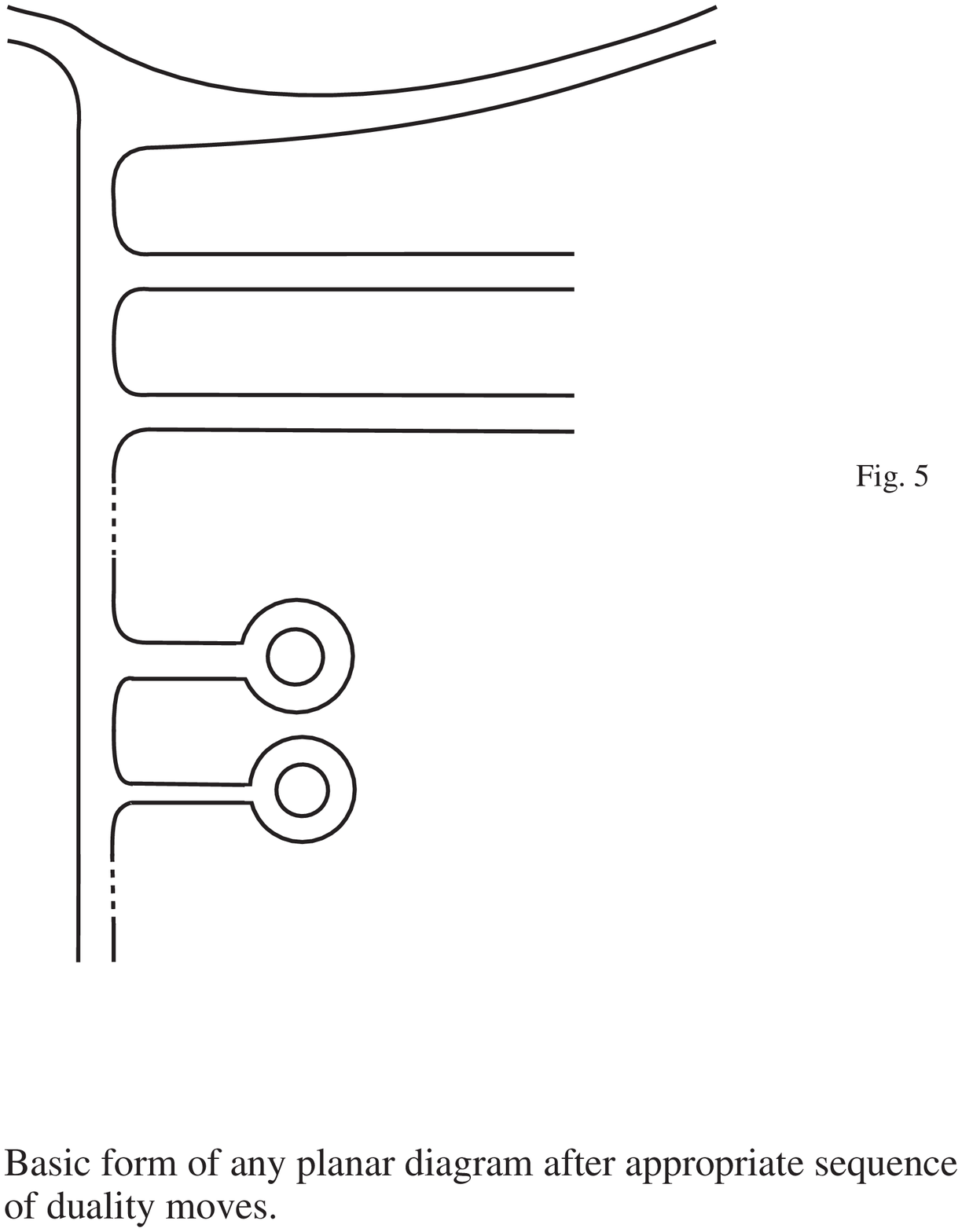}
\end{figure}

\begin{figure}[h]
\epsfxsize=150mm
\epsfbox{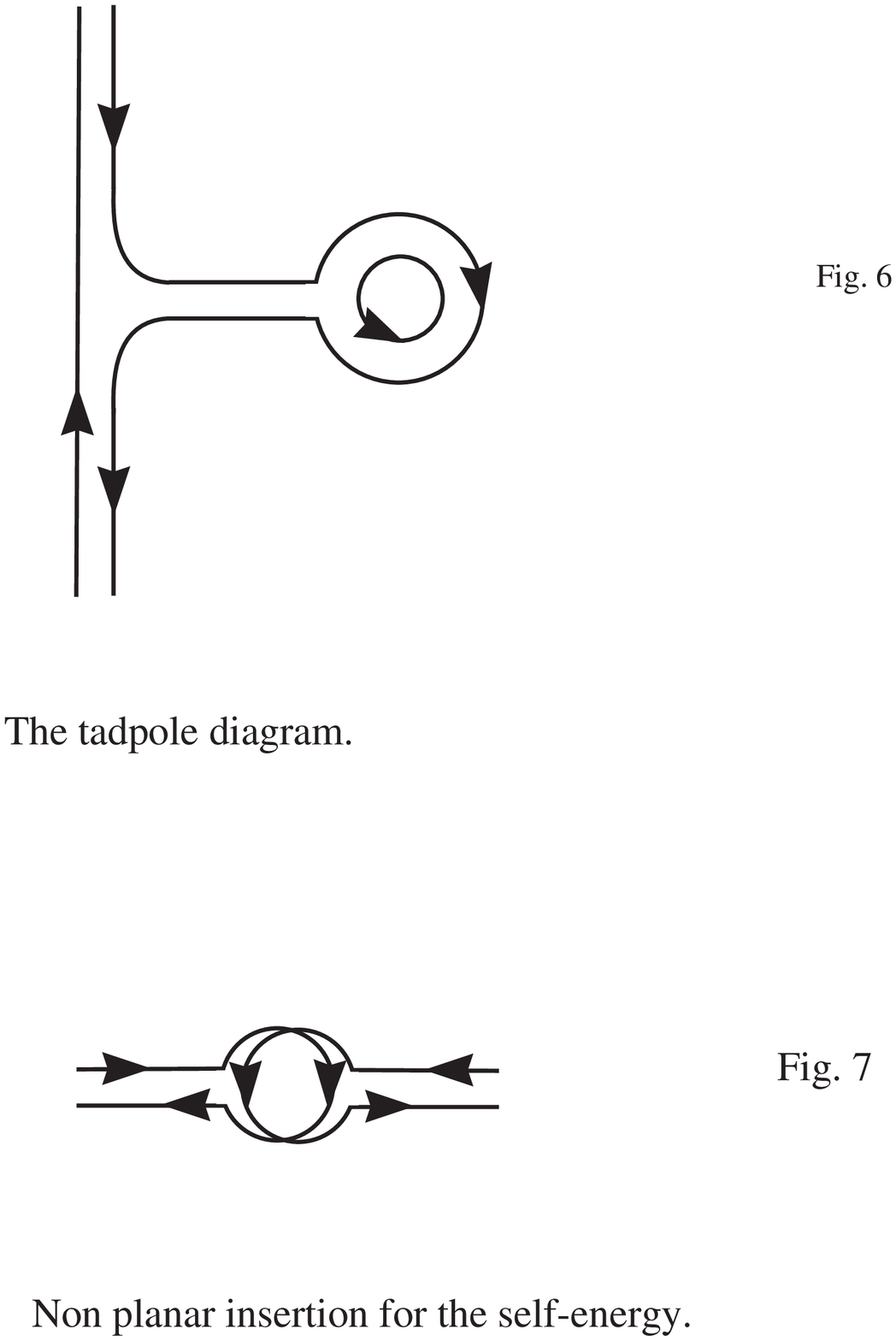}
\end{figure}

\end{document}

%% file: tfdef.tex
\newcommand  {\Rbar} {{\mbox{\rm$\mbox{I}\!\mbox{R}$}}}

\newcommand {\Qbar}
    {\mathord{\setlength{\unitlength}{1em}
     \begin{picture}(0.6,0.7)(-0.1,0)
        \put(-0.1,0){\rm Q}
        \thicklines
        \put(0.2,0.05){\line(0,1){0.55}}
     \end {picture}}}

\newsavebox{\zzzbar}
\sbox{\zzzbar}
  {\setlength{\unitlength}{0.9em}
  \begin{picture}(0.6,0.7)
  \thinlines
  \put(0,0){\line(1,0){0.6}}
  \put(0,0.75){\line(1,0){0.575}}
  \multiput(0,0)(0.0125,0.025){30}{\rule{0.3pt}{0.3pt}}
  \multiput(0.2,0)(0.0125,0.025){30}{\rule{0.3pt}{0.3pt}}
  \put(0,0.75){\line(0,-1){0.15}}
  \put(0.015,0.75){\line(0,-1){0.1}}
  \put(0.03,0.75){\line(0,-1){0.075}}
  \put(0.045,0.75){\line(0,-1){0.05}}
  \put(0.05,0.75){\line(0,-1){0.025}}
  \put(0.6,0){\line(0,1){0.15}}
  \put(0.585,0){\line(0,1){0.1}}
  \put(0.57,0){\line(0,1){0.075}}
  \put(0.555,0){\line(0,1){0.05}}
  \put(0.55,0){\line(0,1){0.025}}
  \end{picture}}
\newcommand{\Zbar}{\mathord{\!{\usebox{\zzzbar}}}}

\newsavebox{\uuunit}
\sbox{\uuunit}
    {\setlength{\unitlength}{0.825em}
     \begin{picture}(0.6,0.7)
        \thinlines
        \put(0,0){\line(1,0){0.5}}
        \put(0.15,0){\line(0,1){0.7}}
        \put(0.35,0){\line(0,1){0.8}}
       \multiput(0.3,0.8)(-0.04,-0.02){12}{\rule{0.5pt}{0.5pt}}
     \end {picture}}